\documentclass[twocolumn]{aastex63}
\pdfoutput=1

\newcommand\tess{{\it TESS }}

\usepackage{amsmath,graphicx,multirow}
 
\received{}
\revised{}
\accepted{}
\submitjournal{ApJ}

\shorttitle{Fast Yellow Pulsating Supergiants}
\shortauthors{Dorn-Wallenstein et al.}

\begin{document}

\title{Short Term Variability of Evolved Massive Stars with \tess \\
II: A New Class of Cool, Pulsating Supergiants}

\correspondingauthor{Trevor Z. Dorn-Wallenstein}
\email{tzdw@uw.edu}

\author[0000-0003-3601-3180]{Trevor Z. Dorn-Wallenstein}
\affiliation{University of Washington Astronomy Department \\
Physics and Astronomy Building, 3910 15th Ave NE  \\
Seattle, WA 98105, USA} 

\author[0000-0003-2184-1581]{Emily M. Levesque}
\affiliation{University of Washington Astronomy Department \\
Physics and Astronomy Building, 3910 15th Ave NE  \\
Seattle, WA 98105, USA}

\author[0000-0002-5787-138X]{Kathryn F. Neugent}
\affiliation{University of Washington Astronomy Department \\
Physics and Astronomy Building, 3910 15th Ave NE  \\
Seattle, WA 98105, USA}
\affiliation{Lowell Observatory, 1400 W Mars Hill Road, Flagstaff, AZ 86001}

\author[0000-0002-0637-835X]{James R. A. Davenport}
\affiliation{University of Washington Astronomy Department \\
Physics and Astronomy Building, 3910 15th Ave NE  \\
Seattle, WA 98105, USA} 

\author[0000-0003-2528-3409]{Brett M.~Morris}
\affiliation{Center for Space and Habitability, University of Bern, Gesellschaftsstrasse 6, 3012 Bern, Switzerland}

\author[0000-0003-0922-138X]{Keyan Gootkin}
\affiliation{University of Washington Astronomy Department \\
Physics and Astronomy Building, 3910 15th Ave NE  \\
Seattle, WA 98105, USA}

\begin{abstract}

Massive stars briefly pass through the yellow supergiant (YSG) phase as they evolve redward across the HR diagram and expand into red supergiants (RSGs). Higher-mass stars pass through the YSG phase again as they evolve blueward after experiencing significant RSG mass loss. These post-RSG objects offer us a tantalizing glimpse into which stars end their lives as RSGs, and why. One telltale sign of a post-RSG object may be an instability to pulsations, depending on the star's interior structure. Here we report the discovery of five YSGs with pulsation periods faster than 1 day, found in a sample of 76 cool supergiants observed by \tess at two-minute cadence. These pulsating YSGs are concentrated in a HR diagram region not previously associated with pulsations; we conclude that this is a genuine new class of pulsating star, Fast Yellow Pulsating Supergiants (FYPS). For each FYPS, we extract frequencies via iterative prewhitening and conduct a time-frequency analysis. One FYPS has an extracted frequency that is split into a triplet, and the amplitude of that peak is modulated on the same timescale as the frequency spacing of the triplet; neither rotation nor binary effects are likely culprits. We discuss the evolutionary status of FYPS and conclude that they are candidate post-RSGs. All stars in our sample also show the same stochastic low-frequency variability (SLFV) found in hot OB stars and attributed to internal gravity waves. Finally, we find four $\alpha$ Cygni variables in our sample, of which three are newly discovered.

\end{abstract}

\section{Introduction} \label{sec:intro}

The exact evolutionary pathway a star takes in its late phases and its final fate as a function of initial mass and composition are both incredibly sensitive to the physics of mass loss. Recent advancements have explored the important roles that interior mixing, pulsations, rotation, binary interactions, and magnetic fields can play on the evolution of the most massive stars \citep{levesque17iau}. However, a discrepancy still exists between the highest mass observed red supergiants (RSGs) and the highest mass observed supernova II-P progenitors (known as the {\it red supergiant problem}, see \citealt{smartt09}). Many solutions have been proposed by observers and theorists alike, including considering extinction effects to account for underestimated progenitor luminosities (e.g. \citealt{walmswell12}, \citealt{kochanek12}); reexamining bolometric corrections that are extremely sensitive to effective temperature to accurately estimate supernova progenitor luminosities(e.g. \citealt{levesque05,levesque06}); attemping to quantify biases in progenitor mass estimates (e.g. \citealt{davies20}, \citealt{kochanek20}); intricately mapping the landscape of explodability in stellar models \citep{sukhbold16,sukhbold18,sukhbold20}; and incorporating well-tested prescriptions for RSG mass loss showing that the highest-mass ($\ge$20M$_{\odot}$) RSGs may simply evolve back to the blue side of the H-R diagram before explosion \citep[e.g.][]{ekstrom12,neugent20}.

A more direct solution is to find yellow or blue stars that have likely already experienced a RSG phase \citep{gordon19}, explicitly determining the mass at which stars no longer end their lives as RSGs. Such post-RSGs allow us to place critical observational constraints on which stars {\it do not} simply evolve redward from the main sequence and then explode. Many methods of finding post-RSGs have been attempted. Surface abundance enhancements of CNO-cycle elements are indicative of both envelope loss and convective mixing that extends from the envelope to the core during the RSG phase. Alternately, stars with evidence of past strong mass loss are likely candidate post-RSGs, detected either via infrared excesses caused by warm circumstellar dust, or by direct detection of ejected mass \citep[e.g.][]{humphreys02,shenoy16}. This, however, requires that the CSM is detectable. Finally, though stars' first crossing of the HR diagram proceeds relatively unimpeded, a small number of very luminous yellow supergiants are observed to encounter the ``yellow void'' where their envelopes become dynamically unstable, resulting in outbursts \citep{nieuwenhuijzen95,stothers01}. 

A possibility that has only recently been explored is to search for stable pulsations in evolved massive stars. Many main sequence massive stars are known pulsators \citep[see, e.g.][]{balona11,blomme11,buysschaert15,johnston17,daszynska18}. Hot stars also exhibit stochastic low-frequency variability (SLFV, manifested as red-noise in the periodogram) that may be attributable to internal gravity waves (IGW; see \citealt{bowman19}, with caveats in \citealt{lecoanet19}). However, beyond lower-mass massive stars found in the upper Cepheid instability strip, $\kappa$-mechanism pulsations are not expected as massive stars evolve redward. $\alpha$ Cygni variables are B and A supergiants (blue supergiants, BSGs) that exhibit microvariability and line profile variations, identified as strange-mode pulsations by \citet{saio13}. These strange modes arise in stars with high ratios of luminosity to mass ($L/M$). \citet{saio13} proposed that $\alpha$ Cygni variables achieve such high $L/M$ values after significant mass loss in the RSG phase. However, while the predicted frequencies roughly correspond to those that are observed, the predicted and observed surface abundances do not match up (though the discrepancy may be resolved by adopting different criteria for convective instability; see \citealt{georgy14}).

While $\alpha$ Cygni variables are promising candidate post-RSG objects, stars evolving from RSGs to BSGs can experience a variety of evolutionary scenarios, including binary interactions or the aforementioned dynamical instabilities, as they evolve blueward. Therefore, a better route may be to search for pulsating {\it yellow} supergiants (YSGs) that are closer to the RSG phase, and which may pulsate for the same reason as $\alpha$ Cygni variables. The discovery of a group of pulsators that are cooler than $\alpha$ Cygni variables, and separate from both the ``yellow void'' and the brightest Cepheids, would thus be of great use. Unfortunately, theoretical modeling of envelope stability in this regime of the HR diagram still encounters convergence difficulties \citep{jeffery16}.

Thankfully, where theory falls short, observations may provide a path forward. The {\it Transiting Exoplanet Survey Satellite} (\tess, \citealt{ricker15}) is collecting lightcurves of the brightest stars across 85\% of the sky. While its primary mission has been to search for exoplanets, many of the brightest massive stars have already been observed at two-minute cadence for approximately 27 days at minimum, with observations of stars in the northern and southern continuous viewing zones (CVZs) lasting an entire year. \tess has thus allowed us to measure microvariability in an unprecedentedly large sample of cool supergiants. In \citet{dornwallenstein19} (Paper I hereafter), we examined a small sample of evolved massive stars that had been observed in \tess Sectors 1 and 2, and found evidence for fast pulsations in three YSGs. Here, we utilize the first 22 Sectors of \tess data, and report the discovery of a group of yellow supergiants that exhibit rapid ($<1$ d) multiperiodic variability. These stars are more luminous and warmer than the classical Cepheid instability strip, fainter than outbursting yellow ``hypergiants,'' and notably cooler than the coolest $\alpha$ Cygni variables. We describe our sample and methodology in \S\ref{sec:data}, then characterize and discuss the SLFV that is ubiquitous in the sample. We present a new class of fast yellow pulsating supergiants in \S\ref{sec:fyps}, discuss the importance of this new class in \S\ref{sec:discussion} before concluding in \S\ref{sec:conclusion}.

\section{Methodology}\label{sec:data}

\subsection{Sample Selection}

We first created a sample of cool supergiants with well-measured effective temperature ($T_{\rm {eff}}$) and luminosity ($L$). \citet{neugent12_ysg} used spectra obtained with the Hydra multi-object spectrograph on the Cerro Tololo 4-meter telescope to confirm the membership of a large sample of YSGs and RSGs in the Large Magellanic Cloud (LMC), along with updated formulae derived from Kurucz \citep{kurucz92} and MARCS \citep{gustafsson08} to obtain $T_{\rm {eff}}$ and $\log L/L_\odot$ from $J-K$ photometry. Because none of the RSGs published in \citet{neugent12_ysg} have been observed by \tess (as described below), we also include the Galactic RSGs from \citet{humphreys78}, \citet{levesque05,levesque06,levesque07}, and the unique RSG WOH G64 \citep{levesque09}. Finally, we discard stars with $\log L/L_\odot \leq 4$ to avoid contamination by lower-mass evolved stars (see, for example, \citealt{levesque17}). 

\subsection{\tess Observations}

We crossmatched our sample of cool supergiants to the latest version of the \tess Input Catalog (TIC, \citealt{stassun18}) available on the Mikulski Archive for Space Telescopes (MAST). From this sample, we selected all stars with a magnitude in the \tess bandpass fainter than $T=4$ (where \tess begins to saturate), and brighter than $T=12$ to obtain sufficient signal-to-noise ratios (SNR) in the lightcurves to detect sub-ppt-level variability (see Paper I). We also omitted any stars with calculated \texttt{contratio} values above 0.1 to mitigate contamination by nearby stars\footnote{See \citet{stassun18} for the exact definition of \texttt{contratio}}. We then downloaded target lists for \tess Sectors 1-22\footnote{\tess target lists are available online at \url{https://tess.mit.edu/observations/target-lists/}}, and select the cool supergiants that have been observed at two-minute cadence. This results in a total of 28 YSGs and 48 RSGs. The positions of the YSGs in the HR diagram have a typical error of 0.015 dex in $\log T_{\rm {eff}}$ and 0.10 dex in $\log L/L_\odot$ respectively.

None of the observed RSGs are in the catalog from \citet{neugent12_ysg}. For RSGs observed by \citet{levesque05,levesque06,levesque07,levesque09}, we used their published estimates of $T_{\rm {eff}}$ and $\log L/L_\odot$ (derived from $M_{bol}$) where available. The $T_{\rm {eff}}$ measurements have typical uncertainties of $\pm25$ K for M stars ($T_{\rm {eff}} \lesssim 3810$ K, $\log T_{\rm {eff}} \lesssim 3.581$), and $\pm100$ K for K stars. The $\log L/L_\odot$ measurements have typical uncertainties of $\sim0.1$ dex. One RSG, V772 Cen (= HD 101712), is a known RSG+B star binary; due to this, the derived $T_{\rm {eff}}$ and $\log L/L_\odot$ from the TIC are both significantly higher than expected for an RSG. A spectrum of V772 Cen is published in \citet{ivanov19}, and available on the {\it Vizier} online service \citep{ochsenbein00}. We obtained this spectrum, and used it to estimate $T_{\rm {eff}}$ and $\log L/L_\odot$ for the RSG member of the binary following \citet{levesque05}, with comparable uncertainties. For the fourteen remaining RSGs, no suitable archival spectrophotometry exists from which we could estimate $T_{\rm {eff}}$ and $\log L/L_\odot$, and so we used the $T_{\rm {eff}}$ estimate published in the TIC, as well as the radius measurement to estimate $\log L/L_\odot$. For stars in both the TIC and \citeauthor{levesque05}, the parameters from the TIC show generally good agreement with the results in \citeauthor{levesque05} to within the errors. However, we do expect the errors on both parameters (especially luminosity) to be significant as the relations used to compute stellar parameters are only validated on dwarfs and giants \citep[see \S2.2 of][]{stassun18}. Table \ref{tab:sample} shows the name, TIC number, coordinates, proper motions, \tess magnitude, $\log T_{\rm {eff}}$, and $\log L/L_\odot$ of each star, as well as the source used to determine their position in the HR diagram, whether the star is a RSG or YSG, and if the star is an $\alpha$ Cygni variable or belongs to the newly identified class of pulsating yellow supergiants (see below). 

\startlongtable
\centerwidetable
\begin{deluxetable*}{lccccccccccc}
\tabletypesize{\scriptsize}
\tablecaption{Names, TIC numbers, coordinates, proper motions, \tess magnitudes and positions in the HR diagram of the cool supergiants observed by {\it TESS,} ordered by effective temperature from coolest to warmest. The source of the $T_{\rm {eff}}$ and $\log L/L_\odot$ measurements is indicated, where N corresponds to \citet{neugent12_ysg}, L to \citet{levesque05,levesque06,levesque07,levesque09}, I to \citet{ivanov19}, and T to the TIC \citep{stassun18}. Typical uncertainties in $\log T_{\rm {eff}}$ and $\log L/L_\odot$ are 0.015 dex and 0.10 dex respectively in \citet{neugent12_ysg}. M stars from \citeauthor{levesque05} have uncertainties of 25 K and 0.1 dex respectively, while the uncertainties in $T_{\rm {eff}}$ in K stars are somewhat larger (100 K). Quantities for RSGs derived from the TIC show good agreement with the values published by \citeauthor{levesque05} where overlap exists. We also indicate whether the star is a RSG or YSG (indicated with ``R'' or ``Y'' respectively), and whether the star is a candidate $\alpha$ Cygni variable or belongs to the newly identified class of pulsators.\label{tab:sample}}
\tablehead{\colhead{Common Name} & \colhead{TIC Number} & \colhead{R.A.} & \colhead{Dec} & \colhead{$\mu_\alpha$} & \colhead{$\mu_\delta$} &  \colhead{$T$} & \colhead{$\log T_{\rm{eff}}$} & \colhead{$\log L/L_\odot$} & \colhead{Source}  & \colhead{RSG/YSG?} & \colhead{Var. Type}\\
\colhead{} & \colhead{} & \colhead{[deg]} & \colhead{[deg]} & \colhead{[mas/yr]} & \colhead{[mas/yr]} & \colhead{[mag]} & \colhead{[K]} & \colhead{$L_\odot$} & \colhead{} & \colhead{} & \colhead{} } 
\startdata
 V1092 Cen & 290678703 & 174.10924243 & -61.31944611 & -6.709 & 0.744 & 5.290 & 3.534 & 4.448 & T & R & - \\ 
 HS Cas & 52782147 & 17.08300080 & 63.58652909 & -2.450 & -0.357 & 5.887 & 3.535 & 4.560 & T & R & - \\ 
HD 143183 & 423407817 & 240.40092730 & -54.14322405 & -2.301 & -3.620 & 4.245 & 3.537 & 5.222 & T & R & - \\ 
BD+35  4077 & 136034302 & 305.30862034 & 35.62126593 & -2.846 & -4.499 & 5.684 & 3.556 & 4.768 & L & R & - \\ 
 AD Per & 348314378 & 35.12084468 & 56.99312317 & -0.066 & -1.423 & 5.357 & 3.543 & 4.587 & T & R & - \\ 
 KY Cyg & 15065085 & 306.49184826 & 38.35213201 & -3.574 & -6.279 & 4.898 & 3.544 & 5.432 & L & R & - \\ 
TYC 8626-2180-1 & 459005094 & 161.46107152 & -59.48870180 & -7.080 & 1.750 & 4.571 & 3.547 & 4.936 & L & R & - \\ 
 V589 Cas & 399355842 & 26.52283837 & 60.99352149 & -0.952 & -0.488 & 5.843 & 3.547 & 4.716 & L & R & - \\ 
 RS Per & 348607532 & 35.60122973 & 57.10947226 & -0.371 & -0.931 & 5.084 & 3.550 & 5.156 & L & R & - \\ 
 V602 Car & 467450857 & 168.37488668 & -60.09134769 & -5.425 & 2.183 & 4.945 & 3.550 & 5.020 & L & R & - \\ 
 W Per & 251118305 & 42.65788594 & 56.98341594 & 0.243 & -1.991 & 5.625 & 3.550 & 4.732 & L & R & - \\ 
 V396 Cen & 443405175 & 199.35433424 & -61.58398415 & -4.770 & -1.758 & 4.580 & 3.550 & 5.212 & L & R & - \\ 
 BI Cyg & 13249363 & 305.34119647 & 36.93214587 & -2.929 & -5.223 & 4.738 & 3.553 & 5.352 & L & R & - \\ 
 BC Cyg & 13325866 & 305.41061705 & 37.53303272 & -3.856 & -5.835 & 5.094 & 3.553 & 5.280 & L & R & - \\ 
 SU Per & 348528265 & 35.52872734 & 56.60413801 & -0.617 & -1.490 & 4.650 & 3.553 & 4.952 & L & R & - \\ 
 PZ Cas & 272324954 & 356.01366443 & 61.78949643 & -3.110 & -1.808 & 4.972 & 3.556 & 5.324 & L & R & - \\ 
 ST Cep & 63963820 & 337.54474090 & 57.00085201 & -3.517 & -2.837 & 5.150 & 3.556 & 4.088 & L & R & - \\ 
 RW Cyg & 15888421 & 307.21079278 & 39.98178278 & -3.255 & -5.511 & 4.596 & 3.556 & 5.156 & L & R & - \\ 
 TZ Cas & 378292562 & 358.23432055 & 61.00233067 & -3.220 & -2.075 & 5.562 & 3.556 & 4.988 & L & R & - \\ 
 BU Per & 264731552 & 34.72204574 & 57.42132329 & -0.526 & -1.106 & 5.898 & 3.556 & 4.764 & L & R & - \\ 
 V349 Car & 457427613 & 157.39738942 & -57.96638247 & -7.191 & 3.632 & 5.250 & 3.559 & 4.808 & L & R & - \\ 
 V774 Cas & 399433806 & 26.75004525 & 60.37232574 & -1.068 & -0.601 & 5.853 & 3.559 & 4.616 & L & R & - \\ 
HD  95687 & 466289471 & 165.39899669 & -61.04883831 & -6.746 & 1.084 & 4.647 & 3.559 & 4.948 & L & R & - \\ 
 V441 Per & 445664243 & 36.34108308 & 57.43726049 & -0.254 & -1.559 & 5.283 & 3.559 & 4.820 & L & R & - \\ 
HD 303250 & 458834083 & 161.08350153 & -58.06484800 & -6.875 & 2.935 & 5.585 & 3.559 & 4.936 & L & R & - \\ 
 RT Car & 458861722 & 161.19645089 & -59.41336782 & -7.450 & 2.914 & 6.417 & 3.559 & 5.260 & L & R & - \\ 
 V772 Cen & 321656644 & 175.45585098 & -63.41457099 & -5.508 & 1.089 & 5.258 & 3.560 & 4.630 & I & R & - \\ 
HD 101007 & 319508664 & 174.23716722 & -61.18277794 & -6.647 & 0.928 & 4.885 & 3.562 & 4.368 & T & R & - \\ 
 V648 Cas & 450147792 & 42.76645187 & 57.85553435 & -0.184 & -1.252 & 5.912 & 3.562 & 4.900 & L & R & - \\ 
 IX Car & 465185147 & 162.60957843 & -59.98238045 & -6.054 & 2.311 & 4.902 & 3.562 & 5.128 & L & R & - \\ 
 W Cep & 65034243 & 339.11484739 & 58.42609816 & -3.329 & -2.132 & 5.207 & 3.566 & 5.466 & T & R & - \\ 
 V910 Cen & 290681168 & 173.93730736 & -61.57806090 & -6.699 & 0.937 & 5.304 & 3.568 & 4.516 & L & R & - \\ 
 V528 Car & 466325776 & 165.77563786 & -60.91072867 & -7.130 & 1.875 & 4.335 & 3.568 & 4.912 & L & R & - \\ 
 YZ Per & 245588987 & 39.60591607 & 57.04616613 & -0.119 & -1.391 & 5.160 & 3.568 & 4.684 & L & R & - \\ 
 V362 Aur & 285640583 & 81.79257440 & 29.92105466 & -0.678 & -2.892 & 4.886 & 3.568 & 4.620 & L & R & - \\ 
 PR Per & 348442493 & 35.42670692 & 57.86281915 & -0.788 & -1.328 & 5.425 & 3.570 & 4.440 & T & R & - \\ 
 FZ Per & 348314886 & 35.24852231 & 57.15832387 & -0.696 & -1.223 & 5.649 & 3.571 & 4.468 & T & R & - \\ 
 V809 Cas & 265186608 & 349.84905043 & 62.73977569 & -2.257 & -2.004 & 4.117 & 3.574 & 4.472 & L & R & - \\ 
 V439 Per & 348671468 & 35.79610521 & 57.19943969 & -0.308 & -0.920 & 5.803 & 3.580 & 4.420 & L & R & - \\ 
 V605 Cas & 348436054 & 35.09359712 & 59.67136417 & -0.711 & -0.953 & 5.710 & 3.585 & 4.920 & T & R & - \\ 
  41 Gem & 337334476 & 105.06593015 & 16.07900049 & -2.088 & -4.853 & 4.230 & 3.597 & 4.341 & T & R & - \\ 
 RW Cep & 422108142 & 335.77923003 & 55.96322672 & -3.616 & -2.349 & 4.370 & 3.597 & 5.470 & T & R & - \\ 
HD 155603 & 188405014 & 258.61523030 & -39.76665102 & -0.900 & -1.087 & 4.138 & 3.601 & 4.870 & T & R & - \\ 
 NR Vul & 435670188 & 297.54969991 & 24.92338263 & -2.320 & -5.807 & 5.421 & 3.602 & 5.348 & L & R & - \\ 
 QY Pup & 334352580 & 116.91052662 & -15.99068889 & -2.162 & 3.511 & 4.898 & 3.608 & 4.756 & T & R & - \\ 
HD  17958 & 390806332 & 44.10270614 & 64.33244354 & -3.739 & 0.017 & 4.124 & 3.623 & 4.548 & L & R & - \\ 
HD  33299 & 367172191 & 77.64572922 & 30.79754031 & -0.015 & -3.142 & 5.274 & 3.633 & 4.044 & L & R & - \\ 
 AZ Cas & 444831689 & 25.56880634 & 61.42120644 & -2.198 & -0.263 & 7.053 & 3.656 & 4.550 & T & R & - \\ 
SK -67   57 & 40603917 & 77.96699736 & -67.16603943 & 1.552 & 0.179 & 11.736 & 3.656 & 4.519 & N & Y & - \\ 
 HV   883 & 30526897 & 75.03151958 & -68.45001791 & 1.785 & -0.043 & 11.196 & 3.680 & 4.841 & N & Y & - \\ 
HD 269953 & 404850274 & 85.05069622 & -69.66801469 & 1.718 & 0.692 & 9.267 & 3.692 & 5.437 & N & Y & FYPS \\ 
HD 269110 & 40404470 & 77.29420213 & -69.60339017 & 2.081 & 0.252 & 10.038 & 3.750 & 5.251 & N & Y & FYPS \\ 
HD 268687 & 29984014 & 72.73273606 & -69.43125133 & 1.833 & -0.114 & 10.465 & 3.784 & 5.169 & N & Y & FYPS \\ 
HD 269840 & 277108449 & 84.04200662 & -68.92812902 & 1.487 & 0.682 & 10.132 & 3.791 & 5.335 & N & Y & FYPS \\ 
HD 269902 & 277300045 & 84.53992899 & -69.10592146 & 1.707 & 0.628 & 9.790 & 3.793 & 5.352 & N & Y & FYPS \\ 
HD 269331 & 179206253 & 79.50763757 & -69.56049032 & 1.772 & 0.291 & 10.114 & 3.810 & 5.307 & N & Y & - \\ 
RMC 137 & 404768745 & 84.65400862 & -69.08552151 & 1.886 & 0.908 & 11.878 & 3.847 & 4.543 & N & Y & - \\ 
CPD-69   430 & 277172433 & 84.23672142 & -69.27176831 & 1.778 & 0.620 & 11.922 & 3.857 & 4.581 & N & Y & - \\ 
W61 27-27 & 277025859 & 84.01580306 & -69.02503035 & 1.526 & 0.556 & 10.770 & 3.861 & 4.493 & N & Y & - \\ 
HD 269392 & 179376451 & 79.96588265 & -69.88570140 & 1.979 & 0.257 & 11.961 & 3.865 & 4.605 & N & Y & - \\ 
HD 269128 & 40518041 & 77.59495623 & -68.77328288 & 1.862 & 0.284 & 9.189 & 3.872 & 5.134 & N & Y & - \\ 
HD 269700 & 425081475 & 82.96784116 & -68.54412683 & 1.602 & 0.401 & 8.808 & 3.882 & 5.069 & N & Y & - \\ 
HD 270151 & 389749856 & 87.26183616 & -70.04170277 & 1.756 & 0.847 & 10.561 & 3.897 & 4.635 & N & Y & - \\ 
CPD-69   491 & 404852071 & 85.20335494 & -69.28089004 & 1.801 & 0.533 & 10.298 & 3.914 & 4.665 & N & Y & - \\ 
HD 270754 & 294872353 & 71.76854552 & -67.11475533 & 0.754 & 0.830 & 11.191 & 3.915 & 4.927 & N & Y & - \\ 
HD 269655 & 391810734 & 82.65769726 & -68.41088876 & 1.495 & 0.972 & 11.123 & 3.924 & 4.497 & N & Y & - \\ 
HD 269997 & 404933493 & 85.33497338 & -69.08535421 & 1.658 & 0.890 & 9.313 & 3.927 & 4.970 & N & Y & - \\ 
W61  6-77 & 389363675 & 85.56405367 & -69.22241953 & 1.658 & 0.424 & 11.069 & 3.969 & 4.502 & N & Y & - \\ 
HD 269777 & 276864600 & 83.57692409 & -67.30380846 & 1.373 & 0.795 & 11.153 & 3.976 & 5.067 & N & Y & - \\ 
CPD-69   394 & 276936320 & 83.65036911 & -69.76013942 & 1.642 & 0.341 & 10.745 & 3.984 & 4.571 & N & Y & - \\ 
HD 269992 & 404967301 & 85.36531401 & -69.80104224 & 1.969 & 0.735 & 9.256 & 3.990 & 5.096 & N & Y & - \\ 
HD 269786 & 277022505 & 83.76501015 & -69.75056435 & 1.872 & 0.508 & 9.655 & 4.000 & 5.116 & N & Y & - \\ 
HD 269101 & 40343782 & 77.43830593 & -68.76940998 & 1.760 & -0.125 & 10.577 & 4.027 & 4.799 & N & Y & $\alpha$ Cyg \\ 
SK -69   68 & 40515514 & 77.49501756 & -69.11716271 & 2.311 & 0.180 & 11.526 & 4.029 & 4.611 & N & Y & $\alpha$ Cyg \\ 
HD 268798 & 30317301 & 74.28356450 & -68.42008185 & 1.942 & -0.074 & 10.103 & 4.033 & 5.071 & N & Y & $\alpha$ Cyg \\ 
HD 269769 & 276936458 & 83.62856915 & -69.78104781 & 1.783 & 0.526 & 10.700 & 4.037 & 4.714 & N & Y & $\alpha$ Cyg \\ 
\enddata
\end{deluxetable*}

Using the Python package \texttt{astroquery}, we queried MAST and downloaded all available two-minute cadence lightcurves for each target. The data are provided by the \tess Science Processing Operations Center (SPOC), and include two flux measurements as a function of time: a simple aperture photometry measurement (\texttt{SAP\_FLUX}) and flux measurements that have been corrected for systematic trends in the data (\texttt{PDCSAP\_FLUX}). The time at each cadence is the photon arrival time at the solar system barycenter, correcting for the position and movement of the \tess spacecraft. For the following, we used the \texttt{PDCSAP\_FLUX} lightcurves. To stitch together lightcurves from different \tess sectors, we divided each sector's lightcurve by its median flux. 

\begin{figure}[ht!]
\plotone{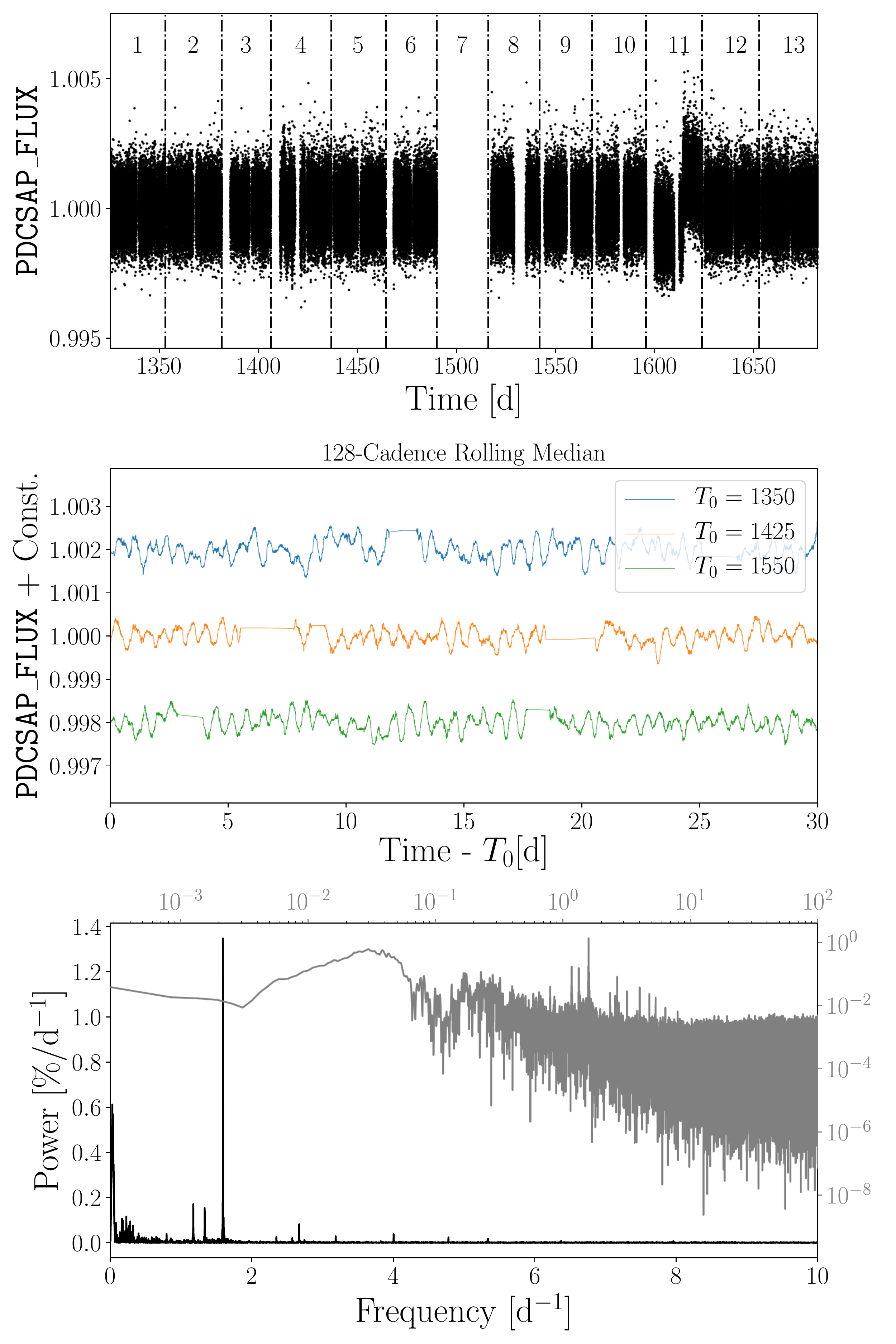}
\caption{({\it Top}): \texttt{PDCSAP\_Flux} lightcurve of the YSG HD 269953. Sector numbers are indicated, with the boundaries between \tess sectors marked as dash-dotted black lines. ({\it Middle}): Same as above, zooming in to three 30-day windows, each beginning at the epoch given in the legend, and smoothing with a 128-cadence rolling median to highlight coherent variability. ({\it Bottom}): Periodogram of the entire unsmoothed lightcurve. Power is multiplied by 100. Linear (logarithmic) scaling for both frequency is shown for the black (grey) line, with corresponding black (grey) axis labels.} \label{fig:example}
\end{figure}

\subsection{Stochastic Low Frequency Variability Across the Upper HR Diagram}

The top panel of Figure \ref{fig:example} shows the lightcurve of the YSG HD 269953, a star that we concluded is a likely post-RSG in Paper I based on its pulsations and infrared excess measured by {\it Spitzer} \citep{bonanos09}. Individual \texttt{PDCSAP\_FLUX} measurements are shown as black points\footnote{A rapid dimming/brightening event is visible in the Sector 11 lightcurve that arises due to a combination of systematics in the detrending and a discontinuity after the mid-sector downlink. However, as revealed in the wavelet analysis below as well as a by-eye inspection of the periodogram computed only on the light curve before this event, this discontinuity only manifests itself as a low frequency transient and has no effect on the recovered frequencies.}. Dash-dotted vertical lines show the boundaries between \tess sectors, and sector numbers are indicated. The middle panel shows a zoom-in to three different thirty-day portions of the lightcurve, now plotting the data after smoothing with a 128-cadence rolling median. Oscillations can clearly be seen. The bottom panel shows the Lomb-Scargle periodogram of the unsmoothed data \citep{lomb76,scargle82}, calculated with the \texttt{astropy} package. We use the \texttt{psd} normalization option, and divide the power by the number of points in the lightcurve. The resulting quantity is equivalent to the absolute value of the power spectral density, $|PSD|$, in units of normalized flux squared. We use the default \texttt{astropy} heuristics for choosing the frequency grid; the maximum frequency is set by the pseudo-Nyquist frequency, $f_{Ny} = 1/(2\langle\Delta t\rangle)$ (where $\langle\Delta t\rangle$ is the mean difference in time between two consecutive observations) and the frequency spacing is five times smaller than the Rayleigh frequency, $f_{R} = 1/T$ where $T$ is the time baseline of the entire lightcurve. The black periodogram is plotted with a linear scaling (corresponding to the black axes labels), and the grey periodogram is plotted with logarithmic scaling on both the frequency and power axes (corresponding to the grey axes labels). Both are scaled to be in units of $\%/d^{-1}$. As found previously in Paper I, the periodogram displays prominent peaks, superimposed upon a frequency-dependent background. The background shows rising power at low frequency that levels off at the lowest frequencies --- i.e., red noise or SLFV -- that is clearly visible in log scaling. 

\begin{figure}[ht!]
\plotone{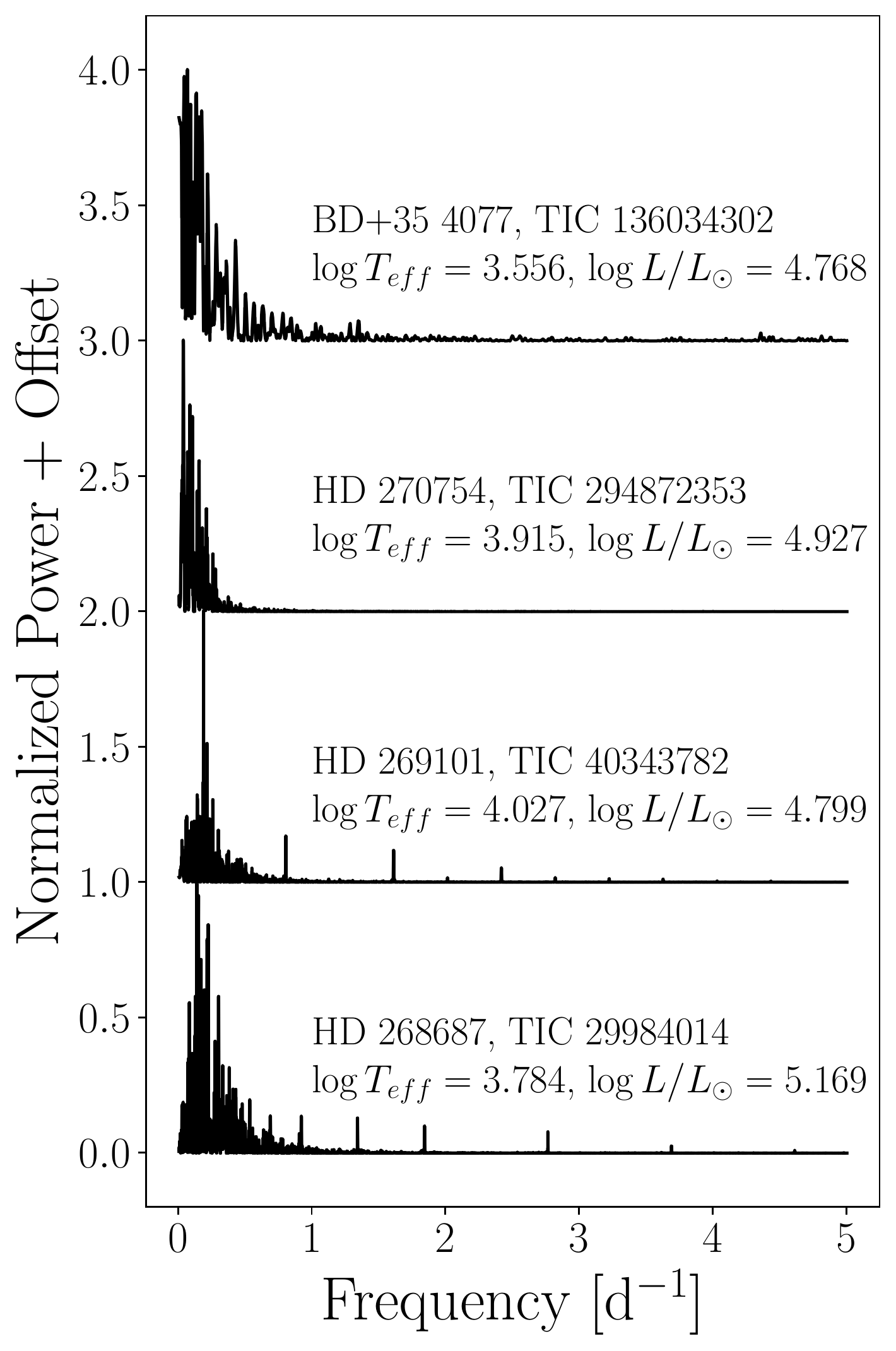}
\caption{Periodograms of four stars that are representative of the entire sample: the RSG BD+35 4077, the YSG HD 270754 (neither of which appear to pulsate), the candidate $\alpha$ Cygni variable HD 269101, and HD 268687, which belongs to the newly identified class of pulsating YSGs. The power has been normalized by the maximum power, and an arbitrary offset has been applied for clarity. Stochastic Low Frequency Variability dominates the power at low frequencies. However, it is possible to see real peaks superimposed on the background of the bottom two periodograms.} \label{fig:rep_periodogram}
\end{figure}

Examining the periodograms of the entire sample shows that SLFV is ubiquitous throughout this region of the HR diagram. SLFV has been identified in hot O and B stars \citep{blomme11,bowman19,bowman19b,bowman20}, and its presence throughout this sample of A-M supergiants suggests that it is in fact a ubiquitous feature of massive stars. Figure \ref{fig:rep_periodogram} shows the periodograms of four stars. The power is normalized to have a maximum value of 1, and an arbitrary offset constant is added for clarity. The top two periodograms are calculated for two ``normal'' supergiants that are representative of the overall sample: the red supergiant BD+35 4077, and the yellow supergiant HD 270754. Their periodograms are dominated by SLFV, and they display no strong peaks. The bottom two periodograms belong to two yellow supergiants: HD 269101 and HD 268687. While both stars' variability are dominated by SLFV, they also show visible peaks in their periodograms.

Searching for significant peaks in a periodogram where the background is independent of frequency (i.e., white noise) is a well-established problem, and peaks can be selected based on various signal-to-noise estimates, or by determining the false alarm probability (FAP), e.g., the probability that a peak with a given power could randomly arise given the null hypothesis of white noise\footnote{Note that this is {\it not} the probability that the peak is not real. In symbols, $P({\rm data}|{\rm noise}) \neq P({\rm noise}|{\rm data})$} \citep{baluev08}. Methods for calculating FAPs in the case of red noise are often incredibly computationally expensive. Fast alternatives have been proposed \citep{delisle20}, however these methods are highly model-dependent, and choosing the wrong noise model can dramatically affect the FAP estimate. Currently, no physical theory for SLFV has been uniformly agreed upon by the community, and so we refrain from utilizing these methods to calculate the FAP.

Instead, we follow \citet{blomme11} and \citet{bowman19,bowman19b,bowman20}, and use the \texttt{curve\_fit} routine within the \texttt{scipy} package to fit the amplitude spectrum ($\alpha(f)$, obtained by taking the square root of the PSD) with a phenomenological model. We adopt the function
\begin{equation}\label{eq:rednoise}
    \alpha(f) = \frac{\alpha_0}{1+(2\pi\tau f)^\gamma}+\alpha_w
\end{equation}
from \citet{stanishev02}, where $f$ is the frequency, $\alpha_0$ is the amplitude as $f\rightarrow0$ in units of normalized flux, $\tau$ is a characteristic timescale in days on which the noise is correlated, $\gamma$ sets the slope of the red noise, and $\alpha_w$ is an additional parameter we add in to model the white noise floor at the highest frequencies, also in units of normalized flux. We note that Equation \eqref{eq:rednoise} is equivalent to the function adopted by \citet{bowman19,bowman19b,bowman20}, and the characteristic frequency in those works is equivalent to $\nu_{char}=(2\pi\tau)^{-1}$. We fit the base-10 logarithm of the amplitude spectrum, calculated as the square root of the PSD, to avoid artificial weighting of real peaks at high frequencies. Note that we do not first prewhiten the coherent variability discussed below from the lightcurves. However, compared to the periodograms calculated from, e.g., the CoRoT lightcurves of hot stars studied by \citet{bowman19b}, the power of the low frequency excess seen in this sample is far stronger than the power of the observed peaks, with the exception of HD 269953, and so we don't expect the fit parameters to be significantly affected. In the case of HD 269953, fitting the logarithm of the amplitude spectrum mitigates any significant effect. To test this, we performed the iterative prewhitening procedure described below, and recorded the values of $\alpha_0$, $\tau$, and $\gamma$ at each step. We found that, after prewhitening the four significant frequencies found below, $\alpha_0$ changed by a factor of 1.4,  $\tau$ by a factor of 1.7, and $\gamma$ by a fact of 0.95. As HD 269953 is the worst case scenario, we decided that these changes do not change the results shown in Figure \ref{fig:fit_params}.

\centerwidetable
\begin{deluxetable*}{lccccc}
\tabletypesize{\scriptsize}
\tablecaption{Names, TIC numbers, fit parameters from Equation \eqref{eq:rednoise}, and corresponding errors for all stars in our sample.\label{tab:fit_params}}
\tablehead{\colhead{Common Name} & \colhead{TIC Number} & \colhead{$\alpha_0$} & \colhead{$\tau$} & \colhead{$\gamma$} & \colhead{$\alpha_w$} \\
\colhead{} & \colhead{}  & \colhead{ppt} & \colhead{d} & \colhead{} & \colhead{ppt} } 
\startdata
 V1092 Cen & 290678703 & $194.9581 \pm 17.3575$ & $0.3464 \pm 0.0318$ & $1.755 \pm 0.049$ & $7.3583 \pm 0.0077$ \\ 
 HS Cas & 52782147 & $414.3034 \pm 33.9049$ & $0.1181 \pm 0.0095$ & $1.784 \pm 0.042$ & $11.6105 \pm 0.0196$ \\ 
HD 143183 & 423407817 & $114.8980 \pm 12.2582$ & $0.2139 \pm 0.0267$ & $1.601 \pm 0.061$ & $7.9003 \pm 0.0111$ \\ 
BD+35  4077 & 136034302 & $198.9679 \pm 25.9046$ & $0.6764 \pm 0.0985$ & $1.544 \pm 0.056$ & $8.5699 \pm 0.0107$ \\ 
 AD Per & 348314378 & $173.5579 \pm 10.9315$ & $0.0597 \pm 0.0039$ & $2.268 \pm 0.092$ & $18.0802 \pm 0.0291$ \\ 
 KY Cyg & 15065085 & $286.2191 \pm 37.0462$ & $0.6387 \pm 0.0998$ & $1.359 \pm 0.042$ & $13.1549 \pm 0.0172$ \\ 
TYC 8626-2180-1 & 459005094 & $546.2201 \pm 69.9013$ & $0.3636 \pm 0.0479$ & $1.456 \pm 0.032$ & $9.0229 \pm 0.0141$ \\ 
 V589 Cas & 399355842 & $92.4473 \pm 7.3775$ & $0.0915 \pm 0.0080$ & $2.142 \pm 0.113$ & $12.1910 \pm 0.0188$ \\ 
 RS Per & 348607532 & $383.0987 \pm 28.7313$ & $0.0859 \pm 0.0069$ & $1.654 \pm 0.037$ & $14.4608 \pm 0.0264$ \\ 
 V602 Car & 467450857 & $163.9396 \pm 10.5384$ & $0.1408 \pm 0.0098$ & $2.165 \pm 0.093$ & $21.4533 \pm 0.0220$ \\ 
\enddata
\tablecomments{This table is published in its entirety in the machine-readable format. A portion is shown here for guidance regarding its form and content.}
\end{deluxetable*}

\begin{figure*}[ht!]
\plotone{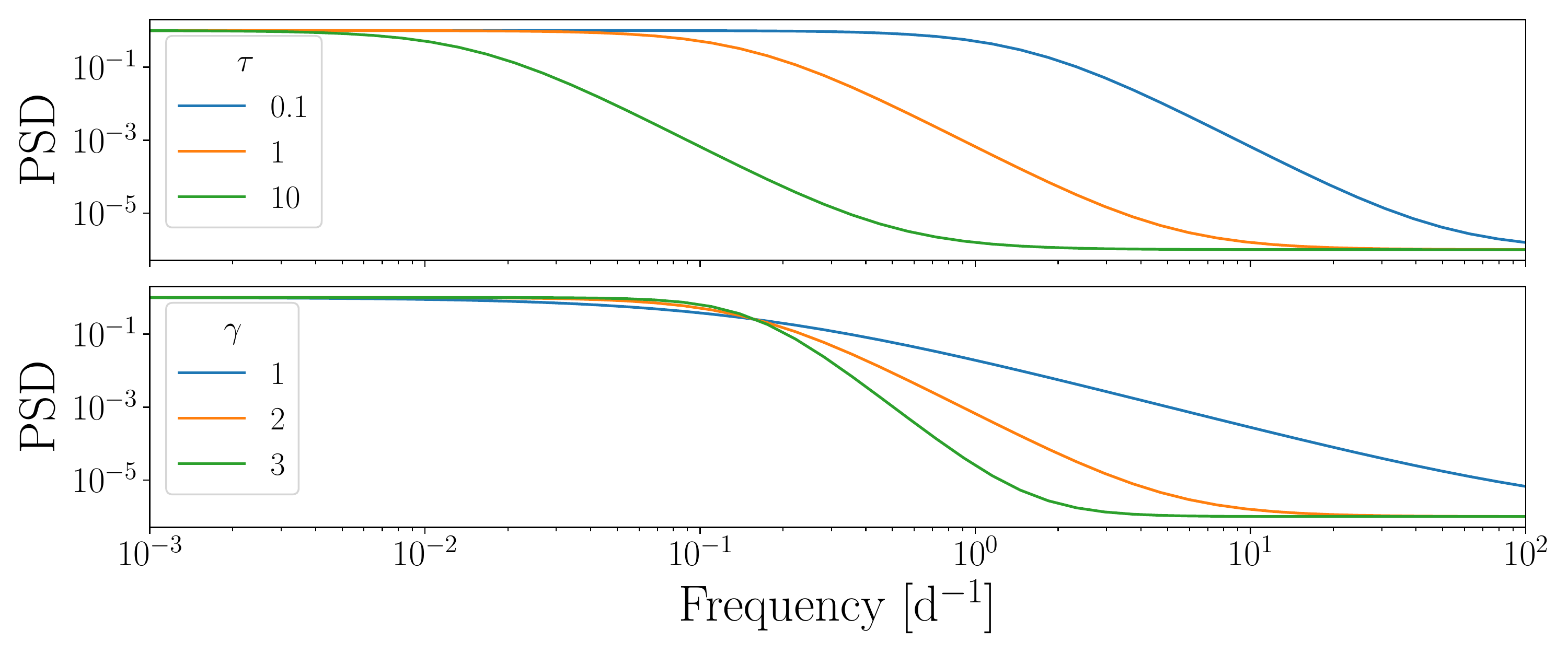}
\caption{Plotting Eq. \eqref{eq:rednoise}, squared so the units are comparable to the Lomb-Scargle periodogram, with $\alpha_0=1$, $\alpha_w=10^{-3}$ and varying $\tau$ (top), and $\gamma$ (bottom).} \label{fig:noise_func}
\end{figure*}

Figure \ref{fig:noise_func} shows the square of Equation \eqref{eq:rednoise}, $|\alpha(f)|^2$, with $\alpha_0=1$, and $\alpha_w=10^{-3}$. The top (bottom) panel shows the effects of varying $\tau$ ($\gamma$) at constant $\gamma$ ($\tau$). Figure \ref{fig:fit_params} shows a summary of our fits to all stars in our sample. The top row shows a histogram of the best-fit values of each of the four fit parameters. We also plot these parameters as a function of $\log T_{\rm {eff}}$ in the bottom row. Errorbars are calculated from the covariance matrix returned by \texttt{curve\_fit}, and the color corresponds to increasing $\log L/L_\odot$ from darker to lighter colors.

\begin{figure*}[ht!]
\plotone{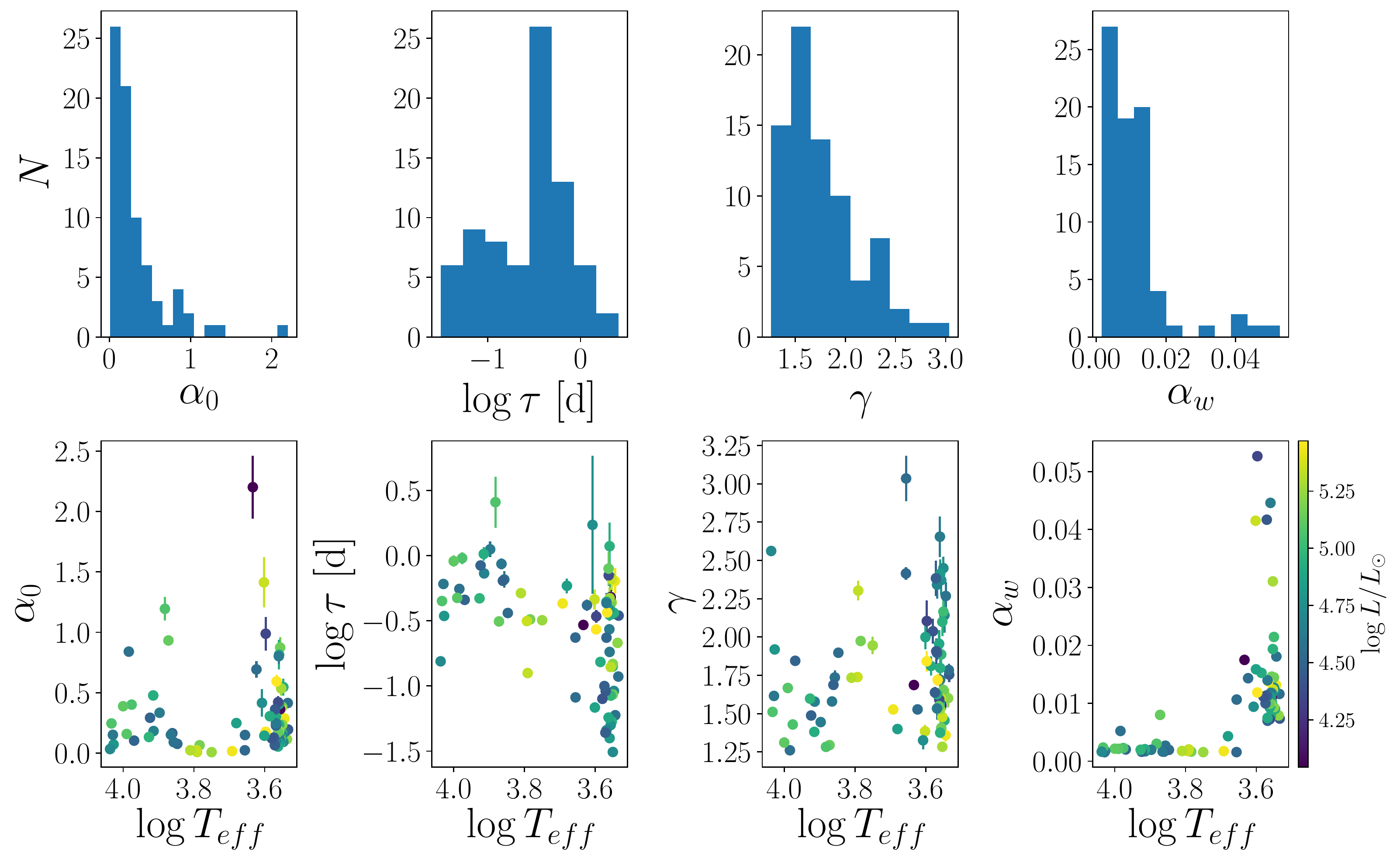}
\caption{Histogram (top) and plot as a function of $\log T_{\rm {eff}}$ (bottom) of the best-fit values for $\alpha_0$ (left), $\tau$ (center-left), $\gamma$ (center-right), and $\alpha_\omega$. Errorbars are calculated from the covariance matrix returned by \texttt{curve\_fit}, and color corresponds with increasing luminosity from purple to yellow.} \label{fig:fit_params}
\end{figure*}

Examining the bottom panels of Figure \ref{fig:fit_params} from left to right: $\alpha_0$, the amplitude of the SLFV, is suppressed at temperatures around $\log T_{\rm {eff}}\sim3.7-3.8$. $\tau$ slightly increases with increasing temperature. While we don't place any bounds on $\tau$ when performing the fits, there is an implicit upper limit to $\tau$ that can be seen in the data, as the SPOC processing pipeline and our sector-combining procedure effectively erases correlations in the lightcurves on timescales longer than a few days. The slope of the noise, $\gamma$, clusters around $\gamma = 2$ for the RSGs, while favoring slightly smaller values for YSGs. Finally, the white noise, $\alpha_w$ is systematically higher for most RSGs. This is not surprising, as all the RSGs in our sample are Galactic, and most were only observed for a single \tess sector. All YSGs are in the LMC, and thus were observed nearly continuously during the first year of \tess observations, with the exception of small gaps when e.g. individual stars passed through the gaps between CCDs in \tess camera 4. Therefore, we expect the signal in the periodogram (relative to the intrumental noise) to be much higher for these stars.

\subsubsection{Convection or Something More?}

As demonstrated above, the characteristic timescale of the background noise, $\tau$, appears to increase with increasing temperatures. Given that $\tau$ roughly corresponds with the characteristic timescale over which the stochastic variability is correlated, this gives us some clues as to the origin of the noise.  Much lower mass yellow stars like our Sun show similar low-frequency power excesses due to granulation in their outer layers. \citet{kallinger14} demonstrate that, from first principles, the characteristic convective timescale, $\tau_{conv}$ (denoted $\tau_{eff}$ in \citealt{kallinger14}) scales with the surface gravity, $g$, and effective temperature as $g^{-0.85}T_{\rm {eff}}^{-0.4}$. Because $g\propto M R^{-2}$ and $R^{-2}\propto T_{\rm {eff}}^4 L^{-1}$, this implies $\tau_{conv} \propto T_{\rm {eff}}^{-3.4}(L/M)^{0.85}$. Thus as a massive star evolves rightward at essentially constant $L$ in the HR diagram, the convective timescale increases strongly as a function of decreasing temperature (with a small boost as the star's $L/M$ increases as the star loses mass). This is the exact opposite of the trend we observe, implying that the low-frequency variability that we see in our sample is not (solely) a result of surface convection, at least in the warmer stars. 

In much hotter massive stars, correlated stochastic variability has been linked with sub-surface convection zones \citep{blomme11}, that may interact with pulsations \citep{perdang09}. More recently, it was suggested by \citet{bowman19} to be a sign of internal gravity waves (IGWs) arising from the boundary of internal convective and radiative layers \citep[though see][for possible caveats]{lecoanet19}. Perhaps the stochastic variability in this sample is connected with that seen in hot stars? Recently, \citet{bowman20} characterized stochastic variability in a sample of 70 OBA stars spanning a range of temperatures above $\sim10^4$ K and masses between $\sim5$ and 80 $M_\odot$ (see Figure 2 in that work). Unfortunately, their sample has very few post-main sequence stars, especially in the mass range of our sample, and so we are unable to construct a complete sequence of $\alpha_0$, $\gamma$, or $\tau \approx 1/\nu_{char}$ as stars evolve across the HR diagram. If stochastic variability is attributable to IGWs, such an evolutionary sequence would be an incredibly powerful means of applying asteroseismology to massive stars as they near the ends of their lives.

It is also possible that sub-surface processes that {\it aren't} IGWs are causing the low frequency stochastic variability. Therefore, we cannot uniquely identify the stochastic variability with any particular source, and reserve such identification for a study of low frequency stochastic variability in massive stars across the entire HR diagram. That said, the RSGs display power law slopes clustered around $\gamma=2$, consistent with what was found by \citet{kiss06} in AAVSO data --- though in a significantly higher frequency range, and with lower amplitude than the lightcurves studied by \citeauthor{kiss06} --- and attributed by those authors to convective processes. Indeed, though the observed scaling of $\tau$ with $T_{\rm eff}$ is inconsistent with surface convection, the timescales of simulated turbulent convection in stellar {\it interiors} (albeit in lower mass stars) do not show this simple scaling \citep{grassitelli15}.

However, \citet{kiss06} found that there is no point at which the power spectra of RSGs turn over; as observing time increases, more power at low frequencies is recovered. Because the \tess observations of RSGs only span a few sectors at most, and correlations on long timescales are smoothed out by the detrending performed by the SPOC, we might expect to see values of $\tau$ clustered at the maximum value possible given the detrending (a few tens of days). However, RSGs display the {\it smallest} observed values of $\tau$ seen in our sample, suggesting that, at the precision of \tess, RSG microvariability is not entirely consumed by convective noise. One possible way to distinguish between subsurface convection and IGWs from core convection may be to measure the macroturbulent velocity of these stars, and compare these measurements with the stars' locations in the HR diagram and observed values of $\alpha_0$, $\tau$, and $\gamma$ (see both \citealt{grassitelli16} and \citealt{bowman20}). It is also entirely possible that both scenarios are at play, and contributing to the observed stochastic variability in this sample.  Regardless of the physical origin, stochastic variability is ubiquitous across the upper HR diagram, from hot stars \citep[e.g.][]{bowman20} to cool supergiants (this work and \citealt{kiss06}). Characterizing this variability and determining its origin has the potential to offer critical insight into the evolution of massive star interiors from birth to death.

\section{Fast Yellow Pulsating Supergiants}\label{sec:fyps}

\begin{figure*}[ht!]
\plotone{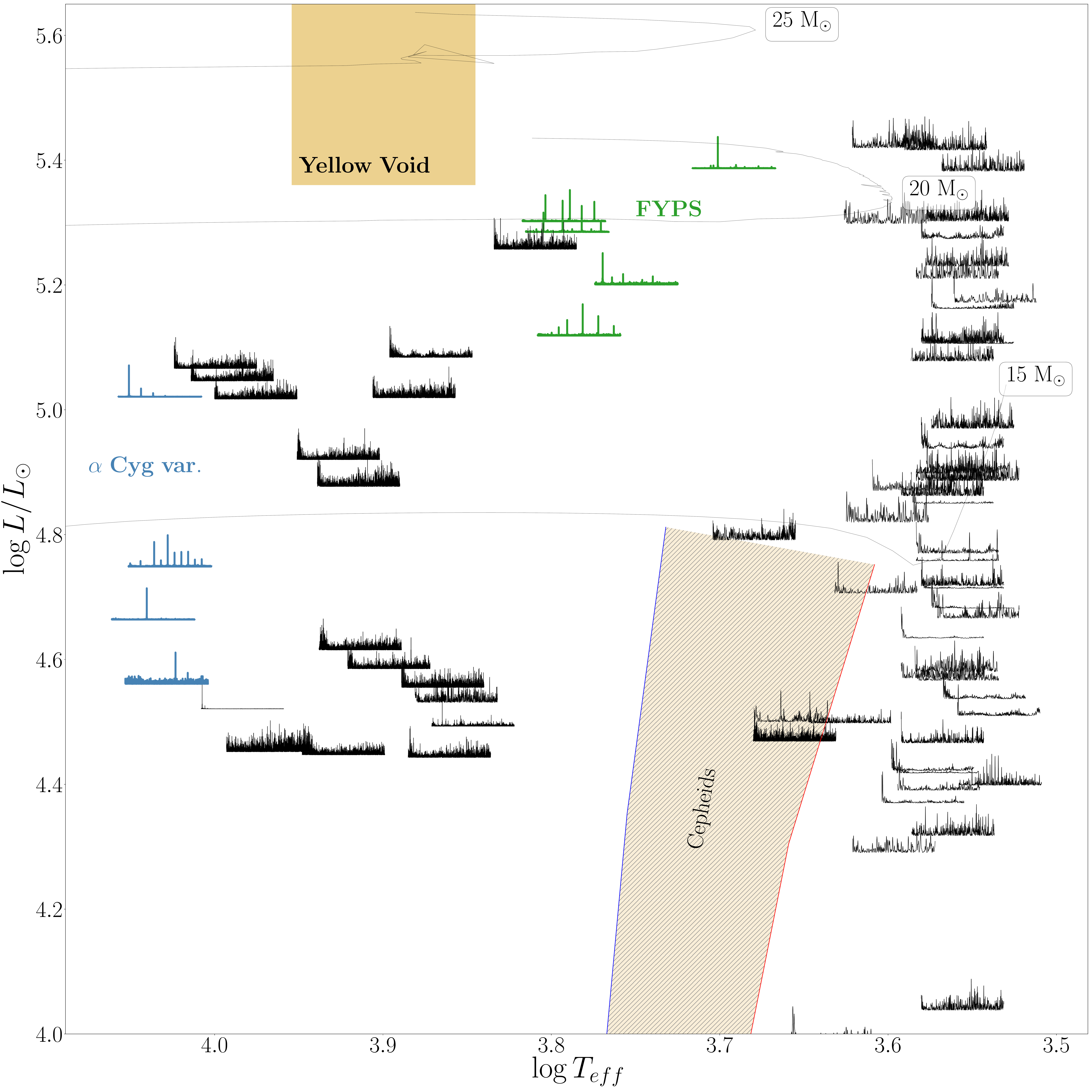}
\caption{HR diagram showing the residual power spectra of each star in our sample between 0 and 5 d$^{-1}$, obtained after dividing out the SLFV in each periodogram, and normalized by the maximum value. Each RPS is centered on the $T_{\rm {eff}}$ and $L$ of the star it corresponds to. The rough boundaries of the yellow void is shown as a goldenrod rectangle. The Cepheid instability strip derived from nonrotating, LMC-metallicity ($Z=0.006$) stellar models on their first crossing of the HR diagram from \citet{anderson16} are shown by the yellow crosshatched region. Four $\alpha$ Cygni variables are highlighted in blue. Separate from all three regions of instability, we find five stars, highlighted in green, that display prominent, high signal-to-noise peaks in their RPS, which we name Fast Yellow Pulsating Supergiants (FYPS). For reference, we plot a subset of the nonrotating, $Z=0.006$ evolutionary tracks calculated with MESA that are described in the text as solid black lines, with their initial masses indicated by the corresponding box.}  \label{fig:dered_hr}
\end{figure*}

After fitting the amplitude spectrum of each star to obtain the best-fit model, $\hat{\alpha}(f)$, we divide its signal out of the power spectrum by computing $|PSD| / \hat{\alpha}(f)^2$. The resulting quantity has no formal definition, but is incredibly useful at showing the power of peaks relative to the background. Hereafter, we refer to the background-normalized power spectrum as the {\it residual power spectrum} (RPS). Figure \ref{fig:dered_hr} shows the HR diagram, where each star in the sample is replaced by its RPS between 0 and 5 d$^{-1}$, normalized by its maximum value and scaled to fit in the plot. Note that because of this scaling, the relative heights of peaks in two different RPS have no relation, but the relative heights of two peaks in the same RPS are meaningful. In particular, plotting the RPS in this way allows us to simultaneously assess the approximate signal to noise in the periodogram as a function of each star's location in the HR diagram. A subset of the nonrotating, $Z=0.006$ evolutionary tracks calculated with MESA and described below are plotted as thin black lines with the initial masses.

For the majority of stars in this sample, their lightcurves appear to be entirely composed of SLFV, and their RPS show either no peaks, or small peaks with low signal to noise. There are, however, two groupings of stars with high signal-to-noise peaks in their RPS. The first group, comprised of a vertical strip of four stars with $\log T_{\rm {eff}} \approx 4.0$ lies in the region of the HR diagram where $\alpha$ Cyg variables are expected to be found\footnote{The frequencies found in the lightcurves of these stars are somewhat higher than in other $\alpha$ Cyg variables \citep{saio13}. However, most observations of $\alpha$ Cyg variables have been taken from the ground, where detecting frequencies around $\sim1$ d$^{-1}$ is difficult. While not the focus of our work, we note these candidate $\alpha$ Cyg variables for use by other authors.}, which we highlight in blue. We searched for any past work that has studied their variability, as listed in SIMBAD \citep{wenger00}, and find the following:

\begin{itemize}
    \item HD 268798 was previously identified as an eclipsing binary by \citet{balona19}, and as a rotational variable with ellipsoidal variations by \citet{pedersen19}. It has not previously been identified as an $\alpha$ Cyg variable.
    \item  HD 269101 was identified as a candidate $\alpha$ Cyg variable by \citet{balona19}, and incorrectly identified as a Slowly Pulsating B-star by \citet{pedersen19} (likely due to its entry in SIMBAD erroneously listing it as an early-B supergiant).
    \item HD 269769 has not previously been studied in the time domain, and thus has not previously been identified as an $\alpha$ Cyg variable.
    \item Sk-69$^\circ$ 68 has not previously been studied in the time domain, and thus has not previously been identified as an $\alpha$ Cyg variable.
\end{itemize}

Apart from these candidate $\alpha$ Cyg variables, we also find a cluster of five stars, all with $5.1 \leq \log L/L\odot \leq 5.5$ and $3.69 \leq \log T_{\rm {eff}} \leq 3.8$ with high signal to noise peaks in their RPS. We highlight these stars in green in Figure 6. This region of the HR diagram contains no other stars\footnote{One star, HD 269331, has a similar luminosity and a $\log T_{\rm {eff}}$ that is 0.02 dex higher than the warmest identified pulsator. Its RPS is low signal-to-noise and shows no significant peaks. In Paper I we identified two prominent bumps in the first two sectors of \tess data. Examination of the remaining sectors shows that these bumps are present throughout the lightcurve, and that the star is displays variability with an amplitude of $\sim1$ ppt. With no significant RPS peaks, we exclude HD 269331 from our subsequent analyses; however it is possible that HD 269331 is a genuine member of this novel class of supergiant pulsator.}. We list the common names (found on SIMBAD), TIC numbers, coordinates, temperatures, and luminosities for the five stars in Table \ref{tab:fyps}. They are well-separated in the HR diagram from the lower-right edge of the yellow void --- a region of the HR diagram occupied by a very small number of stars that exhibit extreme variability and mass loss due to dynamical instabilities in their atmospheres \citep{dejager98} --- which is shown in goldenrod, and the upper-left edge of the Cepheid instability strip (shown in cross-hatched orange, derived from $Z=0.006$, nonrotating stellar models on their first crossing of the HR diagram; see \citealt{anderson16}). This group includes the three pulsating YSGs previously identified in Paper I, as well as two newly identified stars. Many of the frequencies found in their lightcurves (see below) are on timescales shorter than 1 day, and, as discussed in Paper I, are hard to explain with rotational or orbital effects given the large radii of YSGs. Furthermore, while it is possible that these frequencies may arise in the winds of these stars, we deem it unlikely that only YSGs in this region of the HR diagram would show coherent modulations in their winds.\footnote{Regardless, with the notable exception of a small number of incredibly luminous YSGs that have undergone outbursts, YSG winds remain poorly understood.} Finally, spectra of all five stars from \citet{neugent12_ysg} indicate that they are all fairly typical YSGs, though HD 269902 has a slightly weaker \ion{Ca}{2} triplet. Therefore, we adopt the name ``Fast Yellow Pulsating Supergiants'' (FYPS)\footnote{The authors acknowledge the poor adjective ordering in this acronym. However, we believe that FYPS is easier to pronounce than FPYS.} for these stars, and discuss them below.

\begin{deluxetable*}{llllcc}
\tabletypesize{\footnotesize}
\tablecaption{Names, TIC numbers, coordinates, temperatures, and luminosities of the five pulsating YSGs. \label{tab:fyps}}
\tablehead{\colhead{Common Name} & \colhead{TIC \#} & \colhead{R.A.} & \colhead{Dec.} & \colhead{$\log T_{\rm {eff}}$} & \colhead{$\log L/L_\odot$} \\
\colhead{} & \colhead{} & \colhead{$^\circ$} & \colhead{$^\circ$} & \colhead{K} & \colhead{$L_\odot$}} 
\startdata
HD 269953 &	404850274 &	85.050696 &	-69.668015 & 3.692 &	5.437 \\
HD 269110 &	40404470 &	77.294202 &	-69.603390 & 3.750 &	5.251 \\
HD 268687 &	29984014 &	72.732736 &	-69.431251 & 3.784 &	5.169 \\
HD 269840 &	277108449 &	84.042007 &	-68.928129 & 3.791 &	5.335 \\
HD 269902 &	277300045 &	84.539929 &	-69.105921 & 3.793 &	5.352 \\
\enddata
\end{deluxetable*}

\subsection{Chance or New Class?}

All five of the FYPS are located in the LMC. Each \tess pixel is 21'' on a side ($\sim17$ ly at the distance of the LMC). Furthermore, YSGs are found in crowded regions with many hot young stars, making it highly unlikely that the starlight in the optimal aperture defined by the SPOC is coming only from these stars. This effect is somewhat mitigated by \tess's relatively red passband (centered at 7865 \AA); while in bluer passbands, the flux in the aperture may contain significant flux from nearby O and B stars, cool, evolved evolutionary phases of massive stars that dominate the flux in the aperture are significantly rarer due to their shorter lifetimes. Additionally, the binary fraction of massive stars is high \citep{sana12,sana13,moe17}; even the most evolved red supergiants that are the most likely to have interacted and merged with a companion have a binary fraction of $\sim20\%$ \citep{neugent20b}. Therefore it is possible that we are recovering five stars with pulsating companions that happen to be located in a small region of the HR diagram by chance. Assuming the initial massive star binary fraction (as well as the initial period and mass ratio distributions) is roughly constant for the masses of stars in our sample, it is effectively equally likely for any star in our sample to have a pulsating companion (with decreasing likelihood for the largest stars in our sample). Finally, detrending of time series photometry can generate spurious low frequency peaks, which propagate to higher frequencies via harmonics and combinations with real peaks \citep[e.g.][]{tkachenko13}. Potentially, it could appear as if these five stars, which may be ordinary YSGs, are pulsating when they are in fact not.

\begin{figure}[ht!]
\plotone{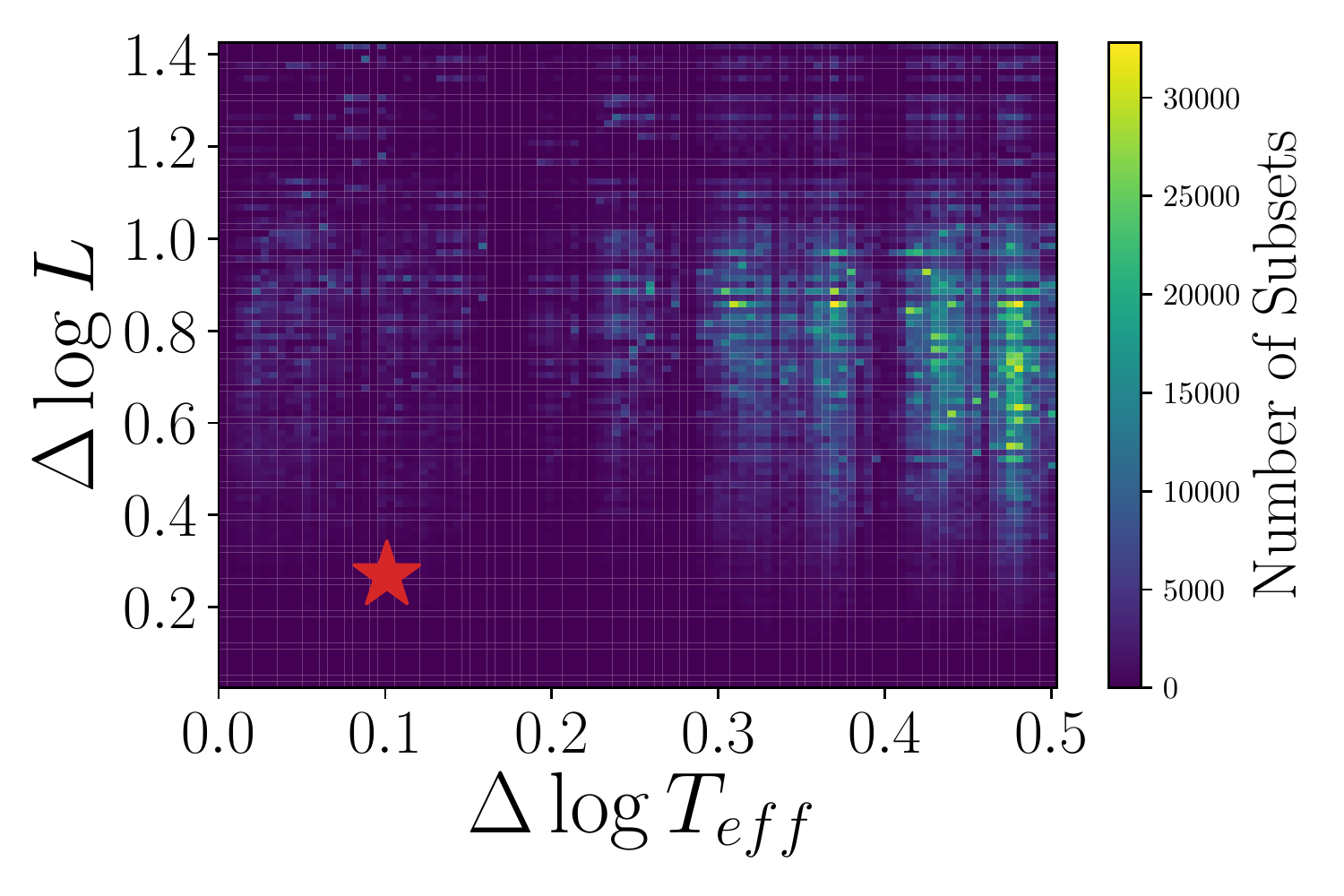}
\caption{Two-dimensional histogram of $\Delta \log T_{\rm {eff}}$ and $\Delta \log L/L_\odot$ for all simulated five-star subsets. The actual $\Delta \log T_{\rm {eff}}$ and $\Delta \log L/L_\odot$ of the real sample of FYPS is shown with the red star. We conclude that the association of FYPS in the HR diagram is unlikely to arise by chance.}  \label{fig:clustering_simulation}
\end{figure}

Therefore, we need to assess the likelihood that the five stars identified above are otherwise normal YSGs, and their lightcurves are all contaminated by starlight from actual pulsating stars (whether from nearby stars in the aperture or a binary companion), or contain spurious periodic signals introduced by detrending. If five stars were randomly selected as ``pulsators'' from our sample due to contamination, we would not expect them to be found in such a small region of the HR diagram. To determine the extent to which crowding may have influenced this detection, we can ask the question: assuming any of the stars in our sample could have randomly been contaminated by pulsators, how likely is it to draw five stars from our sample and have them form a grouping in the HR diagram with equal or lesser size. We can answer this question directly with a simple bootstrap analysis. For each of the 18,474,840 unique subsamples of 5 stars, we calculate the dimensions, $\Delta\log T_{\rm {eff}}$ and $\Delta \log L/L_\odot$, of the smallest box in the HR diagram that contains the subsample. Note that this analysis does not account for the decreasing likelihood of finding a binary companion around a large supergiant, and so in just the case of contamination by a binary companion, the derived likelihood is an uppper limit. Figure \ref{fig:clustering_simulation} shows the two-dimensional histogram of $\Delta\log T_{\rm {eff}}$ and $\Delta \log L/L_\odot$. The actual values of $\Delta\log T_{\rm {eff}}$ and $\Delta \log L/L_\odot$ calculated from the five FYPS is indicated by the red star. We find that only 8.8\% (1.2\%) of all possible subsamples have equal or lesser ranges of $\Delta\log T_{\rm {eff}}$ ($\Delta \log L/L_\odot$). All told, only 0.07\% of all possible subsamples are bounded by a smaller region in the HR diagram. If we repeat this calculation with only the LMC YSGs, this number decreases to 0.03\%. Thus, we deem it exceedingly unlikely that the lightcurves of these five stars (and only these five stars) happened to have randomly been contaminated by a nearby pulsator or pulsating companion, and conclude that we have discovered a genuine new class of pulsating star.

\subsection{Variability and Pulsation Frequencies of FYPS}

\subsubsection{Prewhitening}\label{subsec:prewhitening}

The extraction and measurement of individual pulsation frequencies from lightcurves with frequency-independent noise is a fairly well-defined procedure: the strongest peak can be identified based on its amplitude or FAP, and a prewhitening procedure can be applied to iteratively fit and subtract sinusoids from the lightcurve corresponding to the extracted frequencies until some noise threshold is reached \citep[e.g.,][]{blomme11}. The resulting lists of frequencies, amplitudes, and phases in principle, completely describe the coherent variability found in the lightcurve. However, in the case of frequency dependent noise (in this case, stochastic low frequency variability of astrophysical origin), spurious peaks that are random fluctuations superimposed on the noise are extracted, while true peaks that lie on top of the much lower-amplitude white noise at higher frequencies can be ignored. To account for this effect, we adopt the procedure used by \citet{blomme11}, with the following modifications and stopping criterion:

\begin{enumerate}
    \item At each stage of prewhitening, we fit the (amplitude) spectrum as described above, and obtain the RPS before selecting a frequency to prewhiten. We note that this is the {\it opposite} procedure adopted by \citet{bowman19,bowman19b,bowman20}, who prewhiten coherent frequencies before fitting the amplitude spectra to characterize the stochastic background. This is because the power of the stochastic variability is much higher than that of the peaks in the periodogram, especially at low frequency. However, because we are not removing the stochastic variability from the lightcurve itself, this only helps us locate the peaks in the RPS. The frequencies, amplitudes, and phases we obtain are otherwise identical to what we would obtain following \citet{bowman19,bowman19b,bowman20}.
    \item Paper I identified multiple harmonics of some recovered frequencies. To properly treat potential harmonics, at each stage of prewhitening, we fit both the selected frequency and the amplitude and phase of its first two harmonics. There are some instances where we select a frequency that is itself a harmonic of another lower amplitude frequency, and so the fundamental is not removed by the harmonic fit. We note all instances when this occurs below.
    \item In addition to saving the best-fit parameters of each sinusoid, we calculate the associated errors on each parameter, using the formulae given in \citet{lucy71} and \citet{montgomery99}:
    \begin{align}
        \epsilon(f_j) &= \sqrt{\frac{6}{N}}\frac{1}{\pi T}\frac{\sigma_j}{A_j}\label{eq:ef} \\
        \epsilon(A_j) &= \sqrt{\frac{2}{N}}\sigma_j \\
        \epsilon(\phi_j) &= \sqrt{\frac{2}{N}}\frac{\sigma_j}{A_j}\label{eq:ephi} \\
    \end{align}
    where $N$ is the number of points in the lightcurve, $T$ is the time baseline of the lightcurve, $f_j$, $A_j$, and $\phi_j$ are the frequency, amplitude, and phase extracted at the $j^{\rm th}$ prewhitening stage, and $\sigma_j$ is the standard deviation of the flux at the same stage. We also record the value of the RPS at the selected frequency, and the SNR, calculated as the peak height divided by the standard deviation of the RPS in a narrow window between $f_{max}\pm2 f_R$ and $f_{max}\pm7 f_R$.
    \item As a stopping criterion, we proceed until we reach a minimum in the Bayesian Information Content (BIC, \citealt{schwarz78}) of the fit:
    \begin{align}
        BIC &= -2\ln(\mathcal{L}) + m\ln(N) \\
        -2\ln(\mathcal{L})& = \sum_{i=1}^{N}\frac{(y_i - F(t_i,\Theta_m))^2}{\sigma_i^2}
    \end{align}
    where $\mathcal{L}$ is the likelihood (to within a constant), $m$ is the number of free parameters in the fit ($7j+7$ at the $j^{\rm th}$ stage of prewhitening, beginning with $j=0$), $y_i$ are the original fluxes observed at times $t_i$, $F(t_i,\Theta_m)$ is the sum of all of the fit sinusoids evaluated at fit parameters $\Theta_m$, and $\sigma_i$ are the normalized errors in the original lightcurve \citep[\S 15.1]{press92}.
\end{enumerate}

As a postprocessing step, we discard frequencies that are quite similar to each other (i.e., the difference in frequencies is within $1.5 f_R$, keeping the earliest frequency found). These similar and spurious frequencies can arise due to the short length of the observing baseline \citep{loumos78}. The unique frequencies, amplitudes, and phases, corresponding formal errors, RPS peak heights, and RPS SNRs found for each star are listed in Appendix \ref{app:A}. The frequencies extracted from the FYPS are all between $\sim0.5$ and $\sim4.6$ d$^{-1}$ with semi-amplitudes ranging between $\sim40$ and $\sim280$ ppm.

From the final list of extracted frequencies, we search for harmonics of the form $f_j/f_i = n$, where $n$ is an integer greater than 1, that satisfy
\begin{equation}
        nf_i - f_j \leq \sqrt{\big(n\epsilon(f_i)\big)^2 + \big(\epsilon(f_j)\big)^2}
\end{equation}
 i.e., $f_j$ is an exact integer multiple of $f_i$ to within the errors, and the $k^{\rm th}$ harmonic corresponds to $n = k+1$ (e.g., the first harmonic is $n=2$). We also search for frequency combinations in the form $f_i+f_j=f_k$, such that 
\begin{equation}
        f_1 + f_j - f_k \leq \sqrt{\big(\epsilon(f_i)\big)^2+\big(\epsilon(f_j)\big)^2+\big(\epsilon(f_k)\big)^2}.
\end{equation}

\subsubsection{Wavelet Analysis}

In addition to the frequencies extracted from the entire lightcurve, we also attempt to determine whether the frequencies and amplitudes are stable. To that end, we employ a time-frequency analysis to search for variability in the dominant frequencies. We calculate the Weighted Wavelet Z-transform \citep[WWZ,][]{foster96}, an extension of wavelet analysis with the Morlet wavelet:
\begin{equation}
    \psi(t,\tau,\omega) = e^{i\omega(t-\tau)-c\omega^2(t-\tau)^2}
\end{equation}
where here $\tau$ is the time of the center of the wavelet, $\omega = 2\pi f$ is the angular frequency, and $c$ sets the width of the Gaussian envelope and is chosen to be sufficiently small so that the wavelet decays appreciably over the course of one cycle. Here we adopt $c=0.0125$, following \citep{foster96}; in principle, smaller or larger values of $c$ can be chosen to alter the time and frequency resolution, which is frequency dependent. The discrete wavelet transform can be converted into a projection onto the continuous basis functions
\begin{align}
\Phi_1(t,\omega,\tau) &= 1 \\
\Phi_2(t,\omega,\tau) &= \cos{\omega(t-\tau)}\\
\Phi_3(t,\omega,\tau) &= \sin{\omega(t-\tau)}
\end{align}
and we now fold the Gaussian envelope into a weighting function for each data point at time $t_i$ that depends on the frequency and time center of the wavelet:
\begin{equation}
 w_i(\tau,\omega) = e^{-c\omega^2(t_i-\tau)^2}  
\end{equation}
This change ensures that a wavelet centered on a gap in the data won't pick up small amplitude random fluctuations on either side of the gap, thereby suppressing false power that can often arise in wavelet transformations of unevenly sampled data \citep[e.g.][]{szatmary92,szatmary94}. At each $\tau$ and $\omega$, we calculate the number of {\it effective} data points,
\begin{equation}
    N_{eff}(\tau,\omega) = \frac{\big(\sum_{i} w_i(\tau,\omega)\big)^2}{\big(\sum_i w_i^2(\tau,\omega)\big)}
\end{equation}
and the weighted variance of the flux (which we denote $x$):
\begin{equation}
    V_x(\tau,\omega) = \langle x|x\rangle - \langle \Phi_1 | x\rangle^2
\end{equation}
where the inner product of two functions, $\langle f|g\rangle$ is defined as
\begin{equation}
    \langle f|g \rangle = \frac{\sum_{i}w_i(\tau,\omega) f(t_i)g(t_i)}{\sum_{j}w_j(\tau,\omega)}
\end{equation}
We also calculate the weighted variance of a sinusoidal fit to the model:
\begin{equation}
    V_y(\tau,\omega) = \langle y|y\rangle - \langle \Phi_1 | y\rangle^2
\end{equation}
defining
\begin{equation}
    y = \vec{y}_\Phi\cdot\vec{\Phi}
\end{equation}
where $\vec{\Phi}$ is a vector containing the basis functions, $\Phi$. $\vec{y}_\Phi$ is a vector containing the coefficients of the projection onto the basis functions,
\begin{equation}
    \vec{y}_\Phi = S^{-1}\vec{y}_b
\end{equation}
with the entries of the matrix $S$ equal to $S_{ab}=\langle \Phi_a | \Phi_b\rangle$ and entries in the vector $\vec{y}_b$ are $y_b = \langle \Phi_b | x \rangle$. With these ingredients, we can calculate the WWZ at each $\tau$ and $\omega$ as
\begin{equation}
    WWZ = \frac{(N_{eff} - 3) V_y}{2(V_x-V_y)}
\end{equation}
Finally, we set $WWZ=0$ if $min(t_i-\tau) > 2\pi/\omega$ (i.e., the nearest data point is more than one cycle away from the center of the wavelet), to reduce computation cost.

\subsubsection{HD 269953}

\begin{figure}[ht!]
\plotone{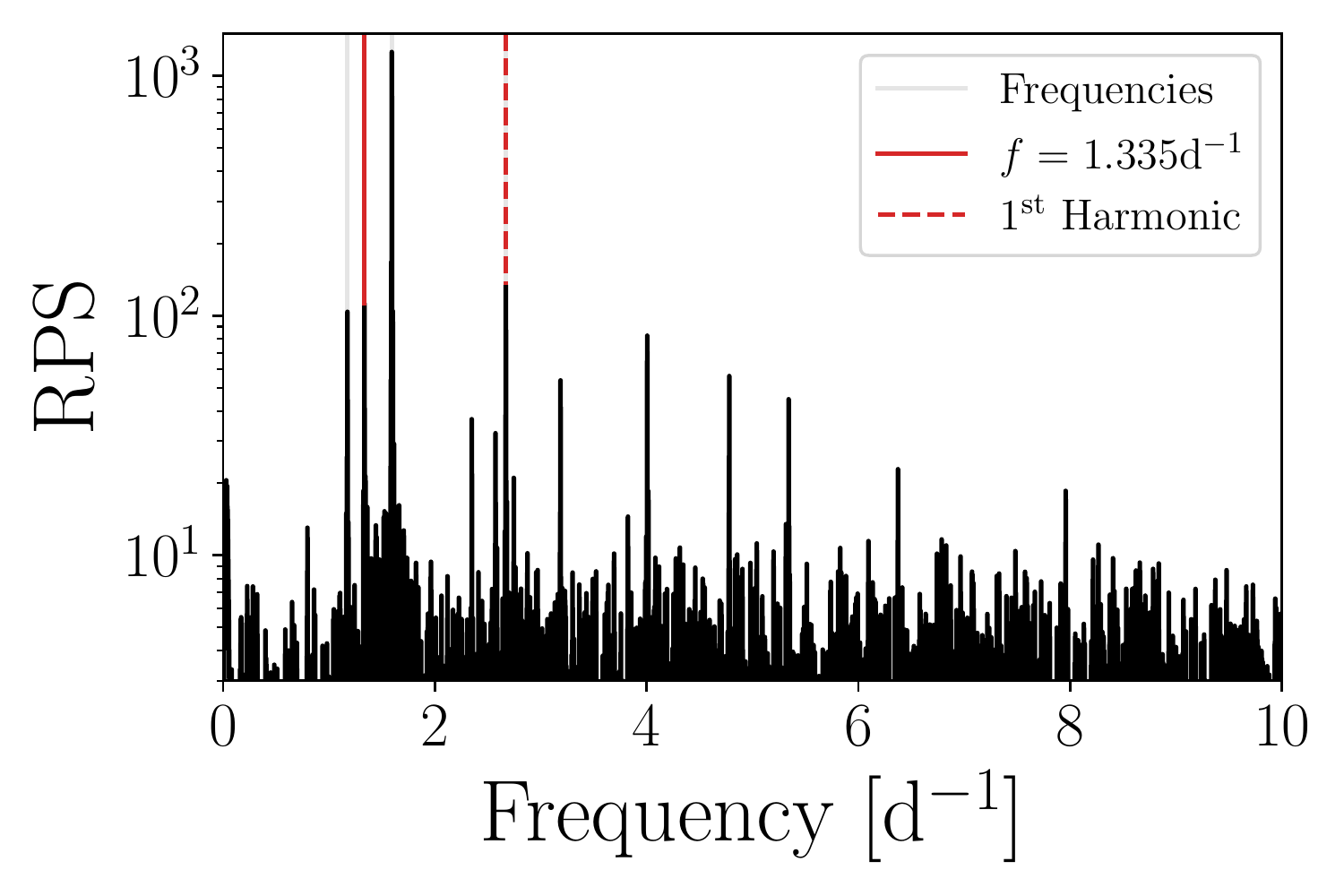}
\caption{RPS for HD 269953, with frequencies extracted by prewhitening indicated by vertical grey lines. Red lines correspond to $f=1.335$ d$^{-1}$ and its harmonic.}  \label{fig:hd269953_rps}
\end{figure}

HD 269953 is the brightest FYPS discovered with $\log L/L_\odot = 5.437$. We initially analyzed it in Paper I, and suggested that it was a post-RSG, largely due to its infrared excess hinting at past mass loss. It was previously studied by \citet{vangenderen06}, who noted its variability. Figure \ref{fig:hd269953_rps} shows the RPS of HD 269953, with the four unique frequencies recovered by prewhitening in grey. After searching for harmonics, we find one instance where the fundamental at 1.335 d$^{-1}$ has a lower amplitude than the first harmonic at 2.671 d$^{-1}$ and so both frequencies are recovered. No combination frequencies are recovered, implying the presence of three independent frequencies in the lightcurve of HD 269953. The presence of these non-aliased frequencies may indicate the presence of multiple oscillation modes in HD 269953. Alternately, the aperture may contain two different pulsating stars. Without a better model of YSG pulsations, we cannot conclusively determine which scenario is more likely; however, if the pulsation mechanism of FYPS is identical to that of $\alpha$ Cygni variables, the presence of multiple modes may not be surprising \citep{kaufer97}.

\begin{figure*}[ht!]
\plotone{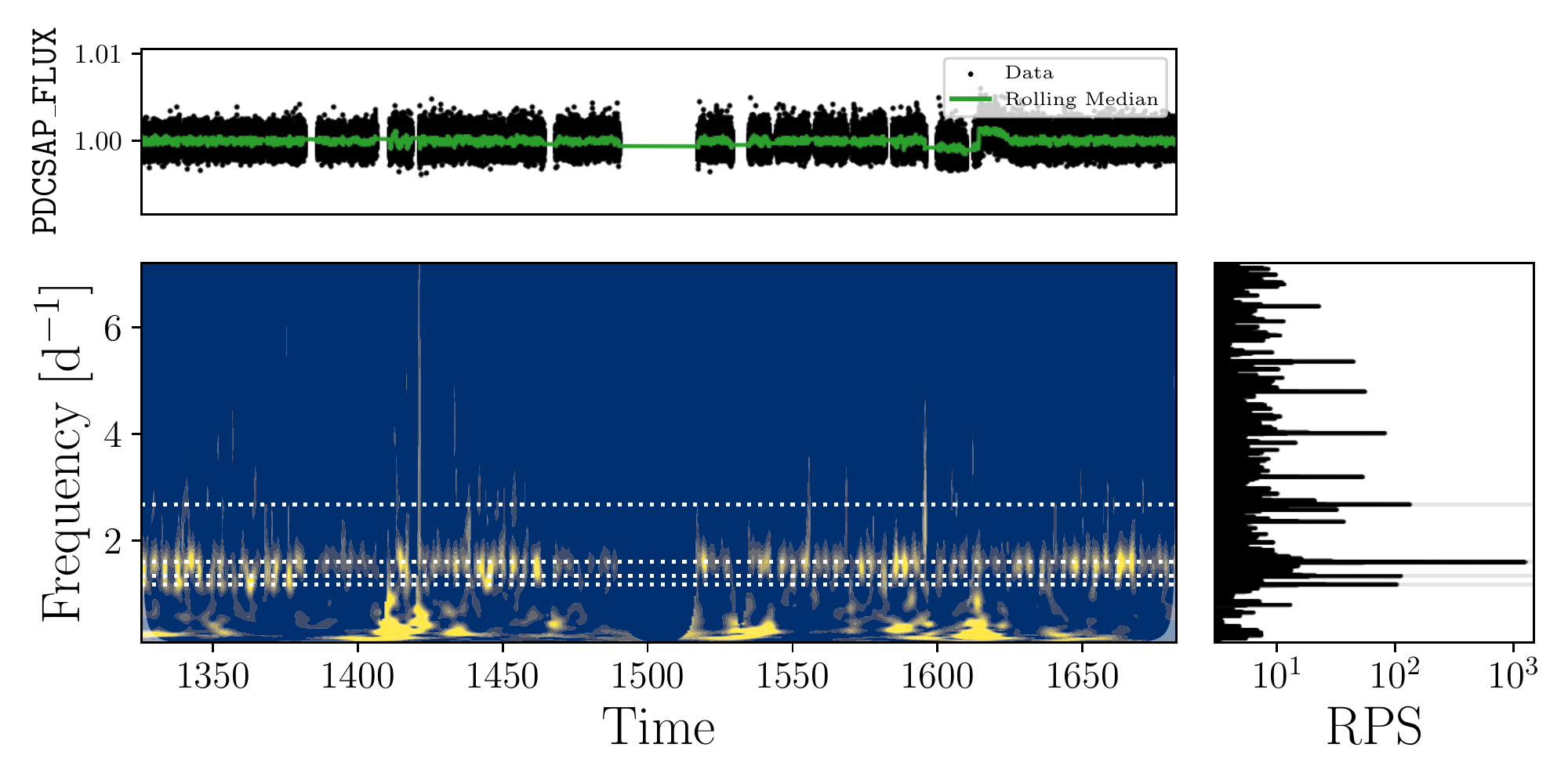}
\caption{({\it Top}): \tess lightcurve of HD 269953 in black points, with green indicating the 128-cadence rolling median. ({\it Right}): RPS, with frequencies identified by prewhitening in grey. ({\it Center}): Weighted Wavelet Z-transform (WWZ, as defined in text) of the lightcurve as a function of time and frequency. Higher values of the WWZ are shown in yellow, and lower values in blue. Identified frequencies are shown as horizontal white lines, and region of the WWZ where the center of the wavelet is within one cycle ($1/f$) of the beginning and end of the data are shaded in white.}  \label{fig:hd269953_wwz}
\end{figure*}

Figure \ref{fig:hd269953_wwz} shows the lightcuve (top, black), rolling median of the flux (top, green), RPS (right), and WWZ (center) calculated on 500 linearly spaced time points, and 1000 frequency points on a $\log_2$ grid between $\log_2(2\pi f) = -1$ ($f\approx0.08$ d$^{-1}$) and 5.5  ($f\approx7.2$ d$^{-1}$). Frequencies extracted via prewhitening are shown as horizontal white (WWZ) or grey (RPS) lines. At frequencies below 1 d$^{-1}$, the WWZ shows transient events that are associated with times where systematics in the data appear to be present --- e.g. the discontinuity after the mid-sector downlink at ${\rm Time}\approx1620$ days. However, the wavelet map demonstrates that these transient events have no effect on the highest amplitude frequencies in the RPS. At higher frequencies, the frequency of maximum power in the WWZ rapidly changes as a function of time, appearing to oscillate between the three lowest extracted frequencies. There even appear to be times when the peak in the WWZ almost disappears (e.g., around ${\rm Time}=1400$; note the gap at ${\rm Time}=1500$ is due to a gap in the data). This indicates that the detected pulsations are not stable, with amplitudes changing on timescales of days. Perhaps the modes are stochastically excited and damped on these timescales. YSG interiors are complex, with multiple boundaries between convective and radiative zones, which may be responsible for driving the pulsations. However, with no reliable models of YSG pulsations, we can only speculate at this time. Of the seven higher frequency peaks detected, five are harmonics of lower-frequency signals as discussed. Some brief, low amplitude transient events are associated with the peaks in the RPS corresponding to harmonics of lower frequency signals. Unfortunately, a drawback of the WWZ (and most time-frequency analyses in general) is that potential interesting high-frequency features are smeared out in exchange for increased time resolution.

\subsubsection{HD 269110}

\begin{figure*}[ht!]
\plotone{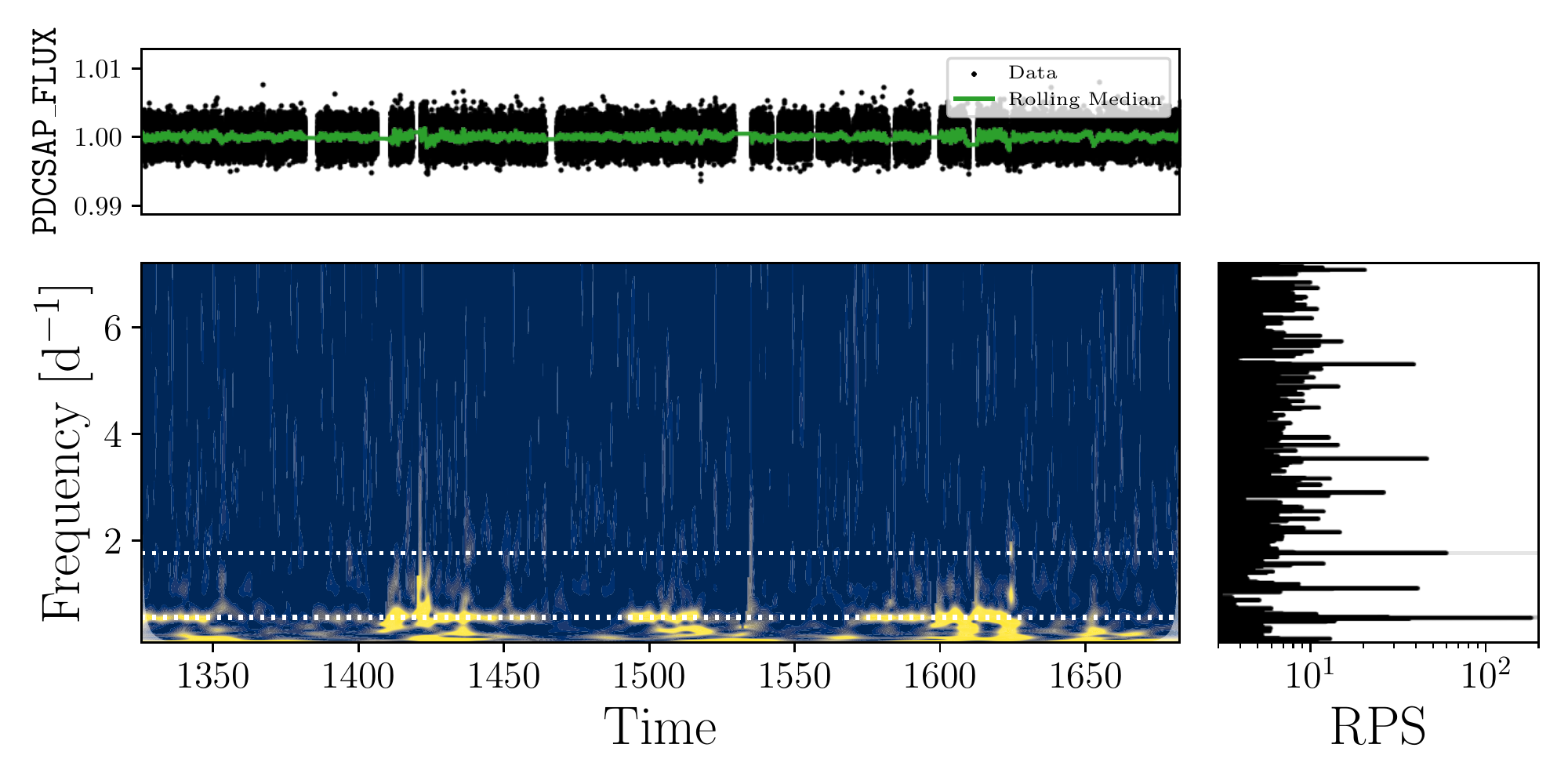}
\caption{Similar to Figure \ref{fig:hd269953_wwz} for HD 269110. Note that the tallest peak in the RPS is the triplet of split frequencies at $f = 0.553$ d$^{-1}$.} \label{fig:hd269110_wwz}
\end{figure*}

HD 269110 was also discussed in Paper I, and had the lowest frequency signals detected there. It has not been analyzed by any other modern variability studies. With our updated prewhitening scheme, we extract four unique frequencies, including three incredibly close to each other: a main peak with the highest SNR at $f = 0.553$ d$^{-1}$, and two small peaks, each separated from the main peak by $\Delta f = \pm 0.011$ d$^{-1}$ ($\sim10$ times the resolution of the periodogram). The peak in the RPS corresponding to the first harmonic of this frequency shows similar structure, though the higher frequency subpeak is very low signal to noise. The lightcurve, RPS, and WWZ are shown in Figure \ref{fig:hd269110_wwz}. Only the three closely spaced peaks can be seen in the WWZ, and those peaks appear to fade and reappear semi-regularly. While the modulation in the WWZ is not sinusoidal, extracting the WWZ in a 0.2 d$^{-1}$ band around the closest frequency to the main peak and calculating the power spectrum reveals a strong peak at $f = 0.010$ d$^{-1}$, quite similar to the frequency difference between the three peaks recovered by prewhitening. One possibility is that HD 269110 is a binary system. Pulsating stars in binaries exhibit frequency modulation similar to what we see \citep[e.g.][]{shibahashi19}, and eccentric close binaries can induce modulations in pulsation amplitudes on the orbital timescale \citep[see][for a lower-mass example]{thompson12}, also in line with what we have detected. An alternate hypothesis is that the $f = 0.553$ d$^{-1}$ is split by rotational effects \citep[\S II.B.3][]{aerts19}, and 0.011 d$^{-1}$ is the rotational frequency. 

If HD 269110 is a binary system with a 20 $M_\odot$ YSG primary and a $1/0.011 \approx 91$ d orbital period, a companion star would have to be $\sim100 M_\odot$ in order for the semimajor axis of the orbit to be larger than the stellar radius derived from the parameters listed in Table \ref{tab:fyps}. Such an object would have to be a black hole in order to not be significantly brighter than the primary, in which case it would still be more massive than any known stellar mass black hole, a scenario we deem to be incredibly unlikely. Alternately, rotational modulation of the WWZ on a $\sim90$ d timescale requires invoking a misalignment of the pulsational and rotational axes. Furthermore, a 91 day rotation period is either quite close to, or exceeds the critical rotation periods of YSGs in the Geneva models, depending on the mass. Regardless, the similarity between $\Delta f$ and the characteristic timescale extracted from the WWZ is intriguing, and warrants follow-up observations. We discuss the implications on the evolutionary status of FYPS below.

\subsubsection{HD 268687, HD 269840, \& HD 269902}\label{subsec:comb}

\begin{figure*}[ht!]
\gridline{\fig{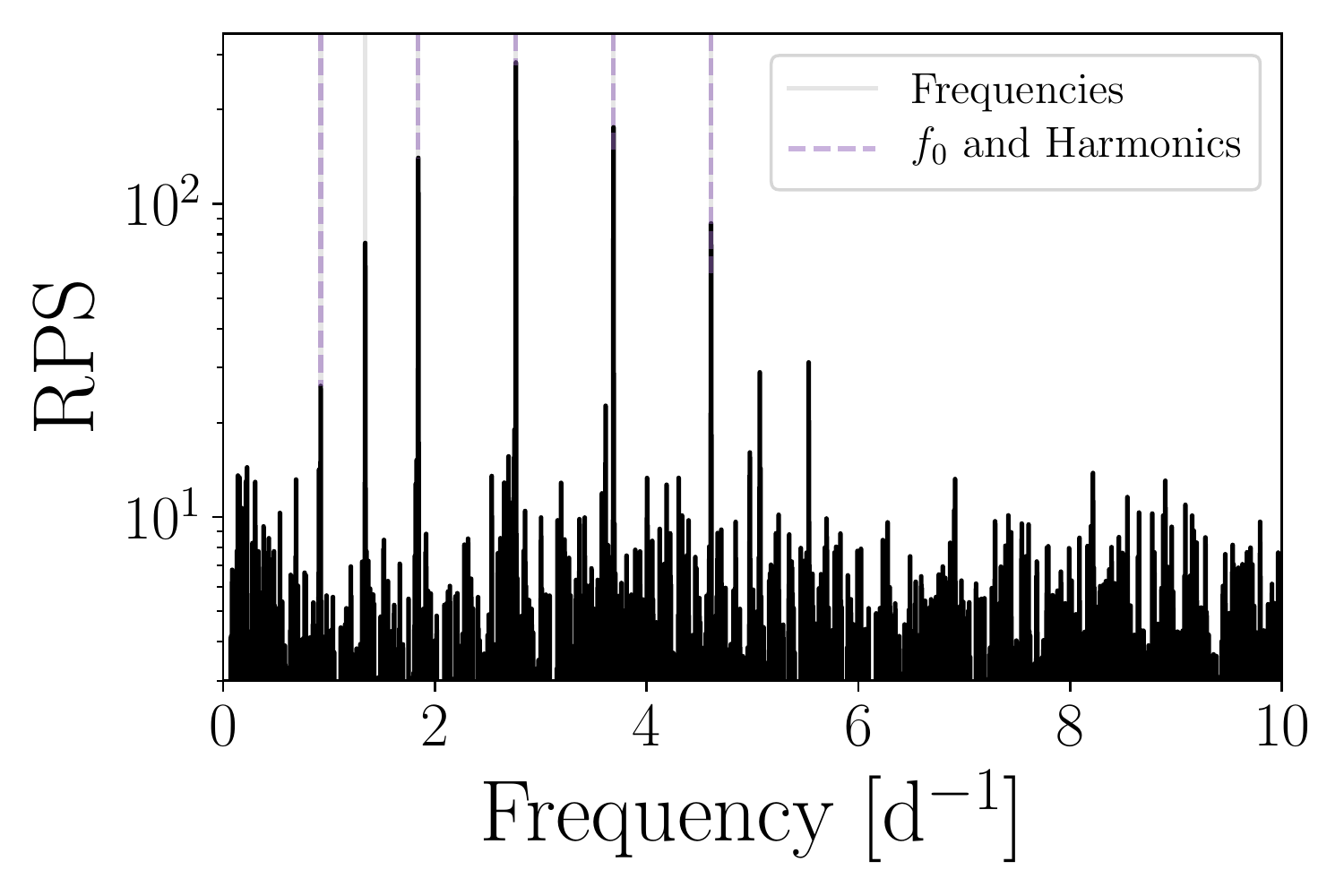}{0.333\textwidth}{}
\fig{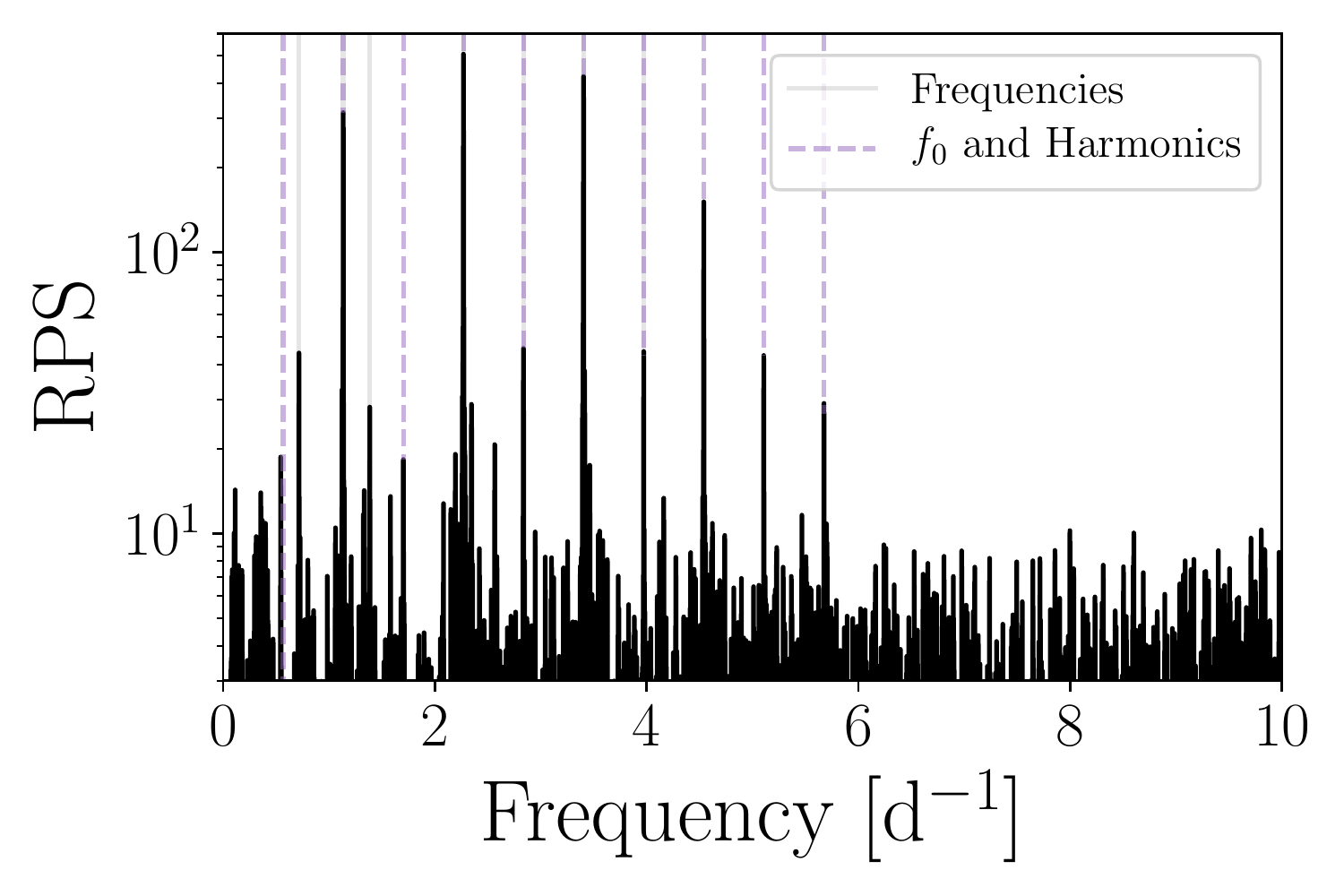}{0.333\textwidth}{}
\fig{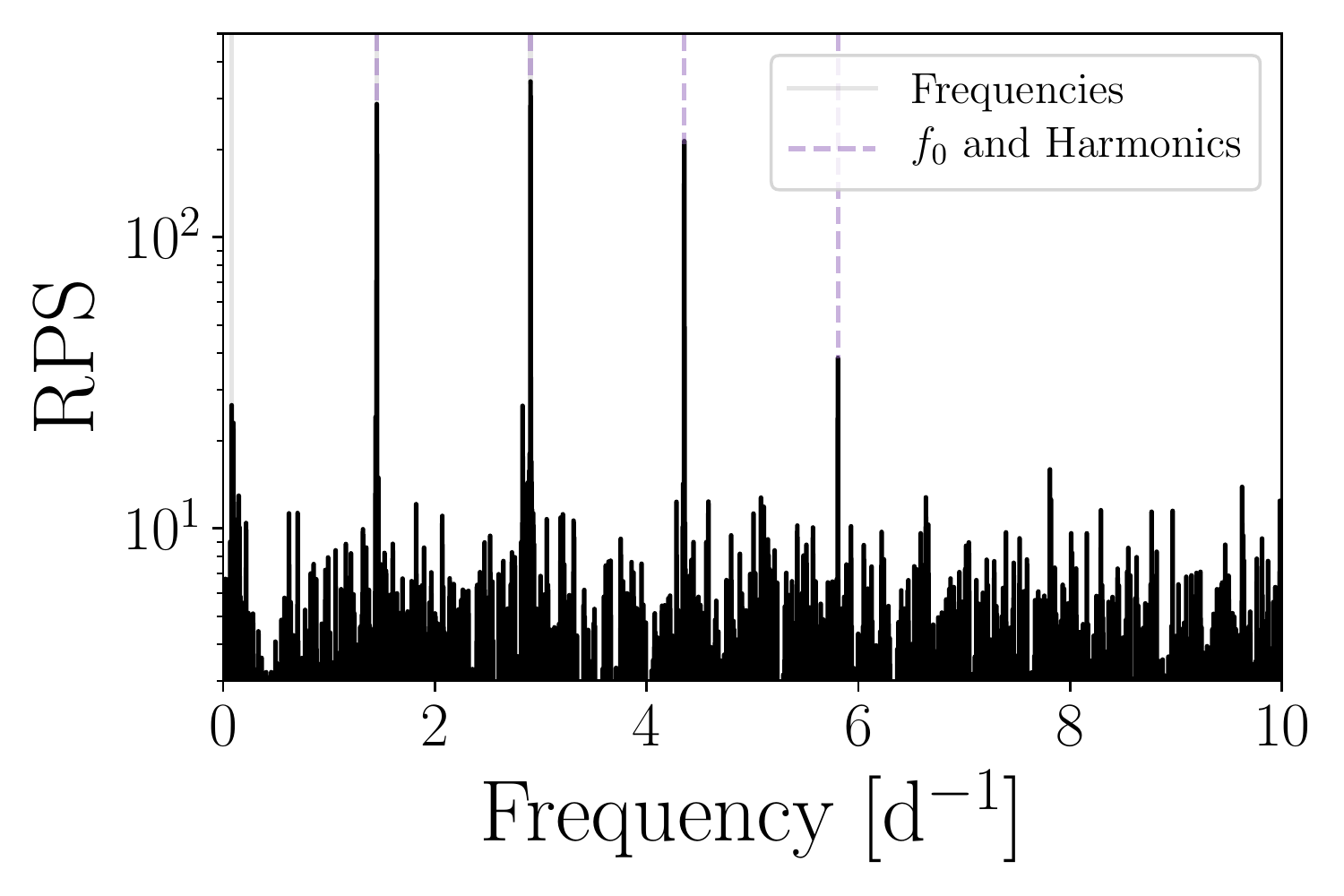}{0.333\textwidth}{}}
\caption{RPS for HD 268687 (left), HD 269840 (center), and HD 269902 (right). Light grey lines show frequencies extracted by prewhitening, and light purple lines show the inferred fundamental, $f_0$, and harmonics, as discussed in text.} \label{fig:comb_rps}
\end{figure*}

HD 268687 was the third candidate pulsating YSG found in Paper I. However, the most prominent frequencies we had found previously were superimposed upon the part of the power spectrum dominated by low frequency stochastic variability. Apart from Paper I, it was identified as a variable star in OGLE photometry \citep{ulaczyk13}. With prewhitening, we extract six unique frequencies. The RPS and extracted frequencies are shown in the left panel of Figure \ref{fig:comb_rps}. The RPS displays a broad comb of peaks, all but one of which correspond to periods faster than one day. None of the low-frequency peaks identified in Paper I are recovered; however, the peaks previously identified as harmonics of the (now nondetected) dominant peak {\it are} recovered. From this, we conclude that the dominant peak identified in early \tess data without correcting for the SLFV may have belonged to the overall pattern of peaks, even though it is no longer detected. Searching for harmonics in the recovered frequencies reveals one frequency at 1.844 d$^{-1}$ with a detected first harmonics at 3.687 d$^{-1}$. A search for combination frequencies shows that the sums of the lowest frequency peak and both the fundamental and first harmonic are also recovered frequencies. Interestingly, the four recovered frequencies in ascending order starting with 1.844 d$^{-1}$ are roughly equally spaced, with a frequency difference of $\Delta f = 0.922$ d$^{-1}$.

Pulsations in both HD 269840 and HD 269902 are newly discovered; the former was identified as a variable in OGLE photometry \citep{ulaczyk13}. Similar to HD 268687, the RPS of both stars show a broad frequency comb (center and right panels of Figure \ref{fig:comb_rps}) with a total of seven and three unique frequencies extracted from HD 269840 and HD 269902 respectively. We do not find any remaining harmonics or combination frequencies amongst the frequencies extracted from both stars. As in HD 268687, the four frequencies recovered in HD 269840 above 2.270 d$^{-1}$ are regularly spaced with a spacing $\Delta f \approx 0.567$ d$^{-1}$ (though the spacing between the highest two frequencies is 0.001 d$^{-1}$ higher). No such regular spacing is found in HD 269902.

Motivated by the apparent regular spacing of peaks in the RPS of all three stars, we searched for evidence that each of the three FYPS exhibits a harmonic chain of peaks. In all three panels of Figure \ref{fig:comb_rps}, we assume that one of the extracted frequencies is a fundamental frequency and plot its first few harmonics. In the case of HD 269840, we instead assume that the frequency of the tallest peak at $f=2.270$ d$^{-1}$ is four times the frequency of the fundamental, which lines up with a peak in the RPS that is not selected by our prewhitening procedure. In all three cases, we find an exceptional match between the assumed harmonic chain and most of the peaks in the RPS. However, the actual observed frequencies are inconsistent with regular harmonic patterns to within the errors

This fact could be due to one of two causes. The offsets from an even spacing pattern could be caused by, e.g., structural glitches \citep[\S IV.B][]{aerts19}. Such glitches can be used to assess sharp features in the stellar structure that would otherwise be inaccessible by other means. Alternately, we may have underestimated the errors for the extracted frequency. Instrumental correlations exist in data taken by the {\it Kepler} and CoRoT missions, and can add to the uncertainty in extracted frequencies \citep{schwarzenberg-czerny03}. As the theory of pulsations in YSGs is still nascent, we have no asteroseismic model for these stars, and thus cannot determine whether the observed offset from an even spacing pattern is astrophysical. However, we can determine the extent to which the uncertainty in the extracted frequencies may be underestimated, assuming that the frequencies {\it should} be evenly spaced.

For each star, we calculate the difference between the observed frequencies and the closest frequency in an evenly-spaced grid extrapolated from the fundamental frequencies assumed above, $\Delta f$, ignoring the observed frequency if the difference between it and the closest predicted frequency is more than 0.005 d$^{-1}$. We also calculated the associated uncertainties, $\sigma_{\Delta f}$ by adding the errors of the observed frequencies and the assumed fundamental frequency in quadrature. Finally, for the entire collection of $\Delta f$ measurements for all three stars, we calculate the reduced chi-squared, 
\begin{equation}
   \chi^2_{red} = \frac{1}{N} \sum \frac{\Delta f}{D \sigma_{\Delta f}} 
\end{equation}
where $D$ is a correction factor to account for underestimated errors, and N is the total number of frequencies in the lightcurves of all three stars. If any scatter in the values of $\Delta f$ around 0 is driven by measurement error, we can find the value of $D$ for which $\chi^2_{red} \approx 1$ --- i.e., the extent to which our errors are underestimated. Doing so yields a value of $D\approx3$, consistent with typical values of $D$ for similar space-based photometric observations \citep[typically $\sim$2-10,][]{schwarzenberg-czerny03}; of course, this requires that we have correctly identified the right fundamental frequency, that the frequency spacing is indeed a regular pattern, and that these offsets are not {\it actually} due to astrophysical effects.

\begin{figure*}[ht!]
\gridline{\fig{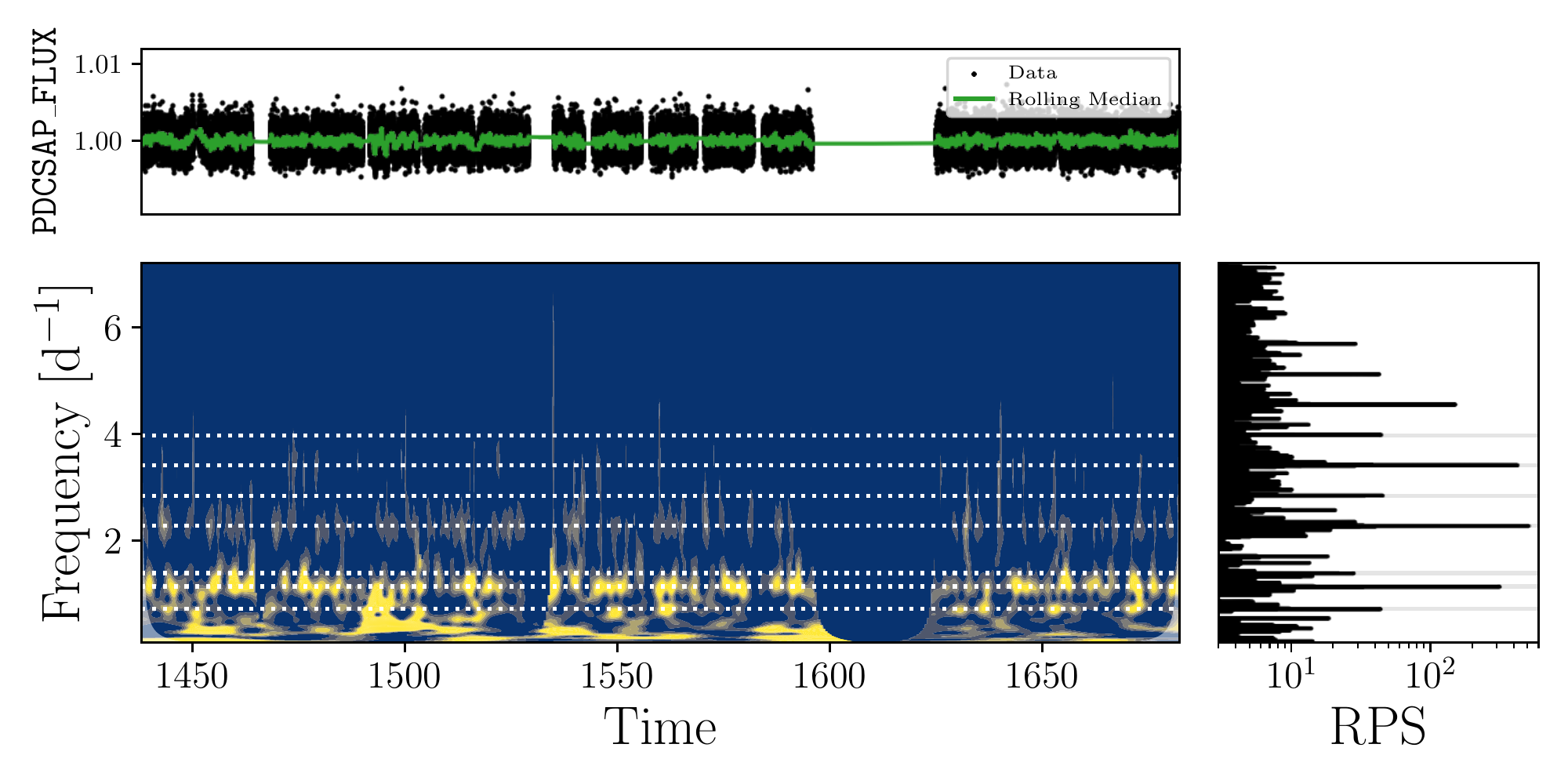}{\textwidth}{}}
\gridline{\fig{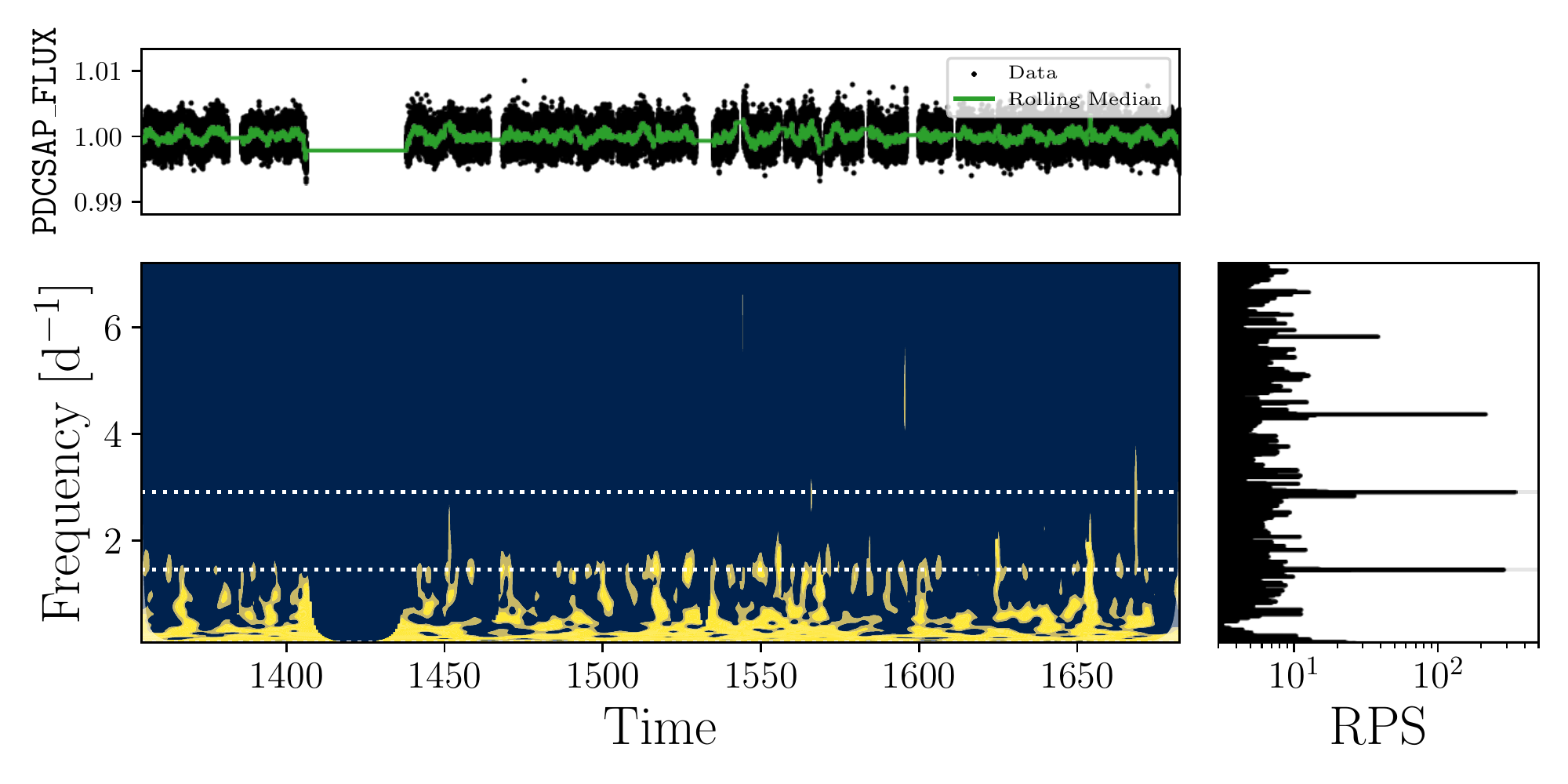}{\textwidth}{}}
\caption{Similar to Figures \ref{fig:hd269953_wwz} and \ref{fig:hd269110_wwz} for HD 269840 (top) and HD 269902 (bottom).} \label{fig:comb_wwz}
\end{figure*}

Ultimately, until systematic correlations in \tess data are better quantified and we are able to generate accurate asteroseismic models of FYPS, we will not be able to make a concrete determination of the cause of the offset between the observed frequencies and a regularly spaced harmonic series. The wavelet analysis used above may be able to help diagnose the behavior of the observed frequencies. Unfortunately, HD 268687 has a measured SLFV amplitude of $\alpha = 0.064\pm0.002$, the highest of all of the discovered FYPS, and the WWZ of the lightcurve is entirely dominated by low-frequency transient features associated with this stochastic variability. The values of $\alpha$ are smaller for HD 269840 and HD 269902. We plot the WWZs of both stars in Figure \ref{fig:comb_wwz}.

In HD 269902, most of the power in the WWZ is contained in the lower frequencies where the stochastic low frequency background dominates --- unsurprisingly given that the star has the second-largest value of $\alpha$ out of the five FYPS ($0.024\pm0.001$). Only by altering the scaling of the image of the WWZ are we able to see the ridge associated with the lowest frequency peak, and it is only detected with low signal-to-noise. In both stars, the the lowest frequency peak in the RPS is associated with a broad band of power in the WWZ, similar to the WWZ of HD 269953, with a secondary band appearing at $\sim2-3$ d$^{-1}$ in HD 269840, at the approximate location of the highest peak in the RPS.

\section{Discussion}\label{sec:discussion}

\subsection{Asteroseismic Modeling}\label{subsec:astero}

The discovery of fast pulsations in YSGs has very interesting implications for the study of the interiors of evolved massive stars. The 25 $M_\odot$ solar-metallicity Geneva model \citep{ekstrom12} has a main sequence lifetime of approximately 7 Myr. It then crosses the HR diagram in under 1 Myr, and half a Myr later, has evolved bluewards once more to become a Wolf-Rayet star \citep{massey17}. Due to the incredibly short lifetime of YSGs on both crossings of the HR diagram, theoretical uncertainty has long stymied our understanding of massive star evolution \citep{kippenhahn90}. Pulsation frequencies extracted from long-baseline lightcurves assembled from space-based observations are perhaps the most precise measurements we can make; typical values of $f/\epsilon(f)$ for frequencies listed in Tables \ref{tab:HD 269953_freqs}-\ref{tab:HD 269902_freqs} are $\sim10^{-4}-10^{-5}$. In better-studied stars, such precision allows for the diagnosis of incredibly complicated physics, and is truly the benchmark of testing stellar evolution theory \citep{aerts19}. No asteroseismic models for YSGs that reliably converge exist \citep{jeffery16}, and so the era of precision YSG asteroseismology is not yet upon us. However, we can compare the frequencies observed in the FYPS with the characteristic Lamb and Brunt-V{\"a}is{\"a}l{\"a} angular frequencies in a model YSG:
\begin{align}
    S_{\ell}^2 &= \frac{\ell(\ell + 1)c_s^2}{r^2} \\
    N^2 &= \frac{g}{H_p}[\delta(\nabla_{ad} - \nabla) + \phi \nabla_\mu]
\end{align}
where $\ell$ is the harmonic degree, $c_s$ is the sound speed, $r$ is the radius of the local shell, $g$ is the gravity, and the remaining variables are defined as in \citet{aerts10}. Modes with angular frequencies $|\omega| > |N|$ and $|\omega| > |S_\ell|$ are mostly restored by pressure ({\it p}-modes), and modes with $|\omega| < |N|$ and $|\omega| < |S_\ell|$ are mostly restored by buoyancy ({\it g}-modes). While rudimentary, such a comparison would illustrate the approximate regions of the stellar structure the observed pulsations will allow us to probe.

\begin{figure*}[ht!]
\plottwo{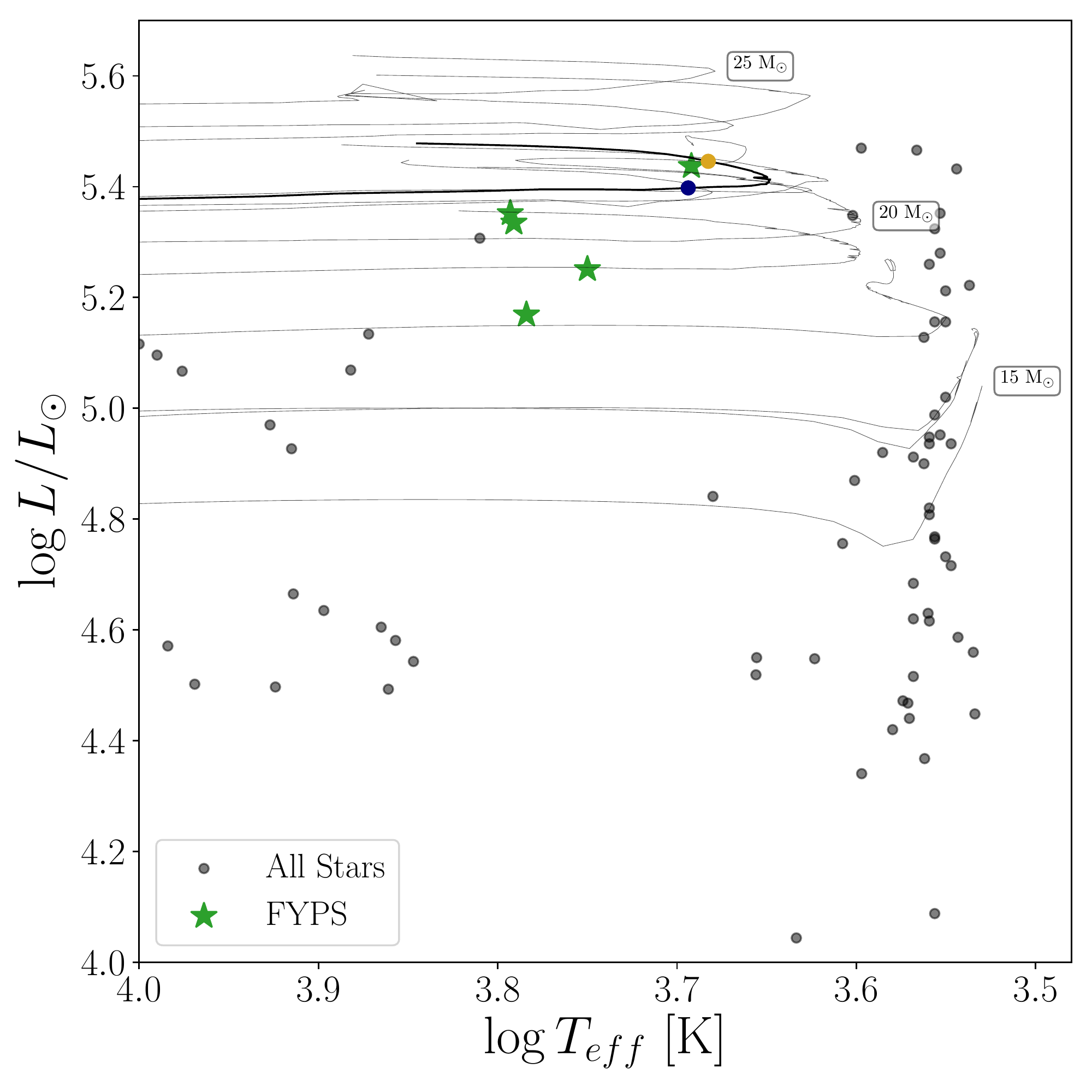}{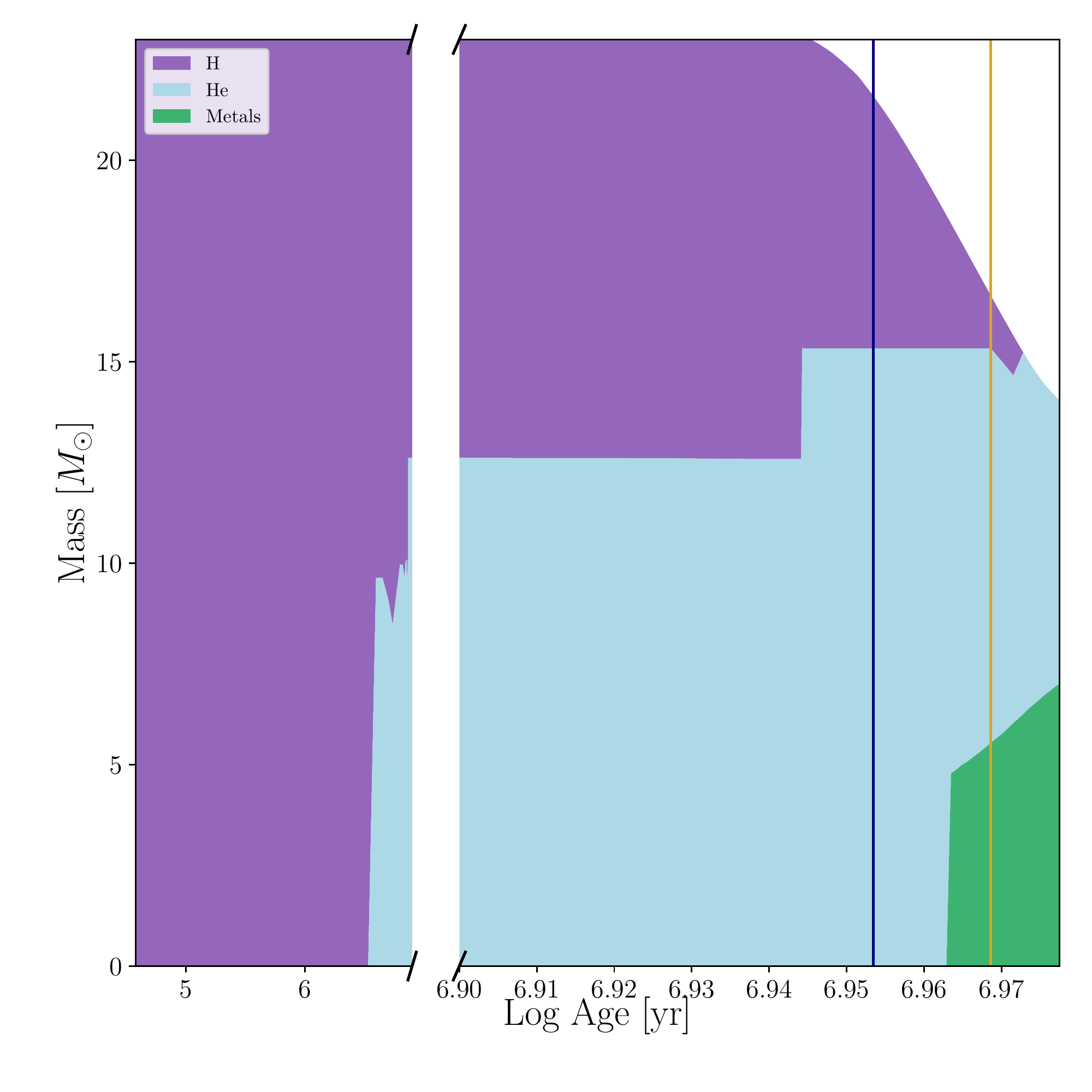}
\caption{{\it Left}: HR diagram with non-FYPS plotted as black points, and FYPS plotted as green stars. $Z=0.006$ evolutionary tracks calculated with MESA are shown as solid grey lines with their initial mass indicated. The 23 $M_\odot$ model is shown as a solid black line. The navy point is the pre-RSG timestep in this model closest to the position of HD 269953 in the HR diagram. The goldenrod point is the closest post-RSG timestep in the same evolutionary track. {\it Right}: Evolution of the structure of the 23 $M_\odot$ MESA model as a function of time. The purple/light blue/green regions are the parts of the star (in mass coordinates) dominated by H/He/metals, respectively. The time axis is broken to illustrate the rapid post-main sequence evolution, including the strong mass loss in the star's final stages. The vertical navy line shows the age of the closest pre-RSG timestep to HD 269953 in the HR diagram. The vertical goldenrod shows the age of the closest post-RSG timestep.} \label{fig:tracks_hr}
\end{figure*}

To do this, we use version 12778 of Modules for Experiments in Stellar Astrophysics (MESA, \citealt{paxton11,paxton13,paxton15,paxton18,paxton19}) to evolve a grid of nonrotating stellar models with initial masses between 15 and 30 $M_\odot$, with a spacing of 1 $M_\odot$, initial metallicity set to $Z=0.006$, and initial helium mass fraction set to $Y=0.25+1.5Z=0.259$. We first evolve the models through the entire pre-main sequence stage until the entirety of the star's luminosity comes from nuclear burning then introduce more elaborate physics. Convective mixing follows the {\tt mlt++} prescription from \citet{paxton13}, with overshoot following \citet{farmer16}. For mass loss, we use the ``Dutch'' \citep{glebbeek09} and ``Vink'' \citep{vink01} cool/hot wind schemes respectively, and adopt an efficiency of $\eta=0.8$ for the former \citep[e.g.][]{maeder01}. Because the luminosity in the outer layers of red supergiants can exceed the Eddington luminosity due to high opacity in these layers, we follow \citet{ekstrom12}, and increase the mass-loss rate in by a factor of 3 when the luminosity exceeds five times the Eddington luminosity. Finally, we use Type 2 opacities for when the star has extra C/O during and after He burning. With these controls, all but the 28 $M_\odot$ model successfully ran; as we focus the remainder of our analysis on a single model, we chose not to introduce additional controls for this one model. Our inlist files, including our timestep and spatial resolution controls, are available online at \url{https://github.com/tzdwi/TESS}.

The post-ZAMS evolutionary tracks are plotted in the left panel of Figure \ref{fig:tracks_hr}, with the 15, 20, 25, and 30 $M_\odot$ tracks labeled. We note that evolutionary modeling of post main-sequence massive stars is fraught with uncertainty; different prescriptions for overshooting, or mass loss \citep[e.g.][]{martins13}, or different choices of input physics like binary interactions or rotation \citep[e.g.][]{dornwallenstein20} can radically alter the evolutionary pathway a given stellar model might take. Nonetheless, we can use these evolutionary tracks to find a model that may approximate the structure and evolution of the FYPS. Interestingly, the stars with initial mass $M \geq 18 M_\odot$ lose enough mass to begin turning around on the HR diagram. Models more massive than 19 $M_\odot$ become luminous and warm enough to encounter the yellow void, at which point the models begin to exhibit rapid changes in their luminosities and effective temperatures on $\sim$month to year timescales --- for clarity, we do not show the post-RSG portion of the tracks after they reach an effective temperature hotter than 7000 K. The stars in our sample are plotted as grey points, and the FYPS are plotted as green stars. The dark line shows the model that passes the closest to the position of HD 269953 in the HR diagram, which has an initial mass of 23 $M_\odot$. We show the evolution of the interior structure of this model throughout its lifetime in the right panel of Figure \ref{fig:tracks_hr}. Each colored region shows the part of the star (in mass coordinates) dominated by H (purple), He (light blue), and metals (green) respectively as a function of the age of the star.

The navy point and navy line in both panels of Figure \ref{fig:tracks_hr} indicate the pre-RSG timestep whose temperature and luminosity best match the observed values for HD 269953. At this point, the model is $\sim9.0$ Myr old, has a current mass of $21.6 M_\odot$, and has $\log T_{\rm {eff}} = 3.694$, and $\log L/L_\odot = 5.398$. By this time, the star has begun core He fusion, has created $\sim0.8 M_\odot$ of C, and is losing significant mass from its envelope. Similarly, the goldenrod point and vertical line in Figure \ref{fig:tracks_hr} correspond to the post-RSG timestep in the same model that is closest to HD 269953. This model is 9.3 Myr old, has undergone extensive mass loss as an RSG, and has $\log T_{\rm {eff}} = 3.683$ and $\log L/L_\odot = 5.446$. It is still fusing He in its core, but has built up significant C and O mass. It's current mass is only 16.6 $M_\odot$

\begin{figure*}[ht!]
\plotone{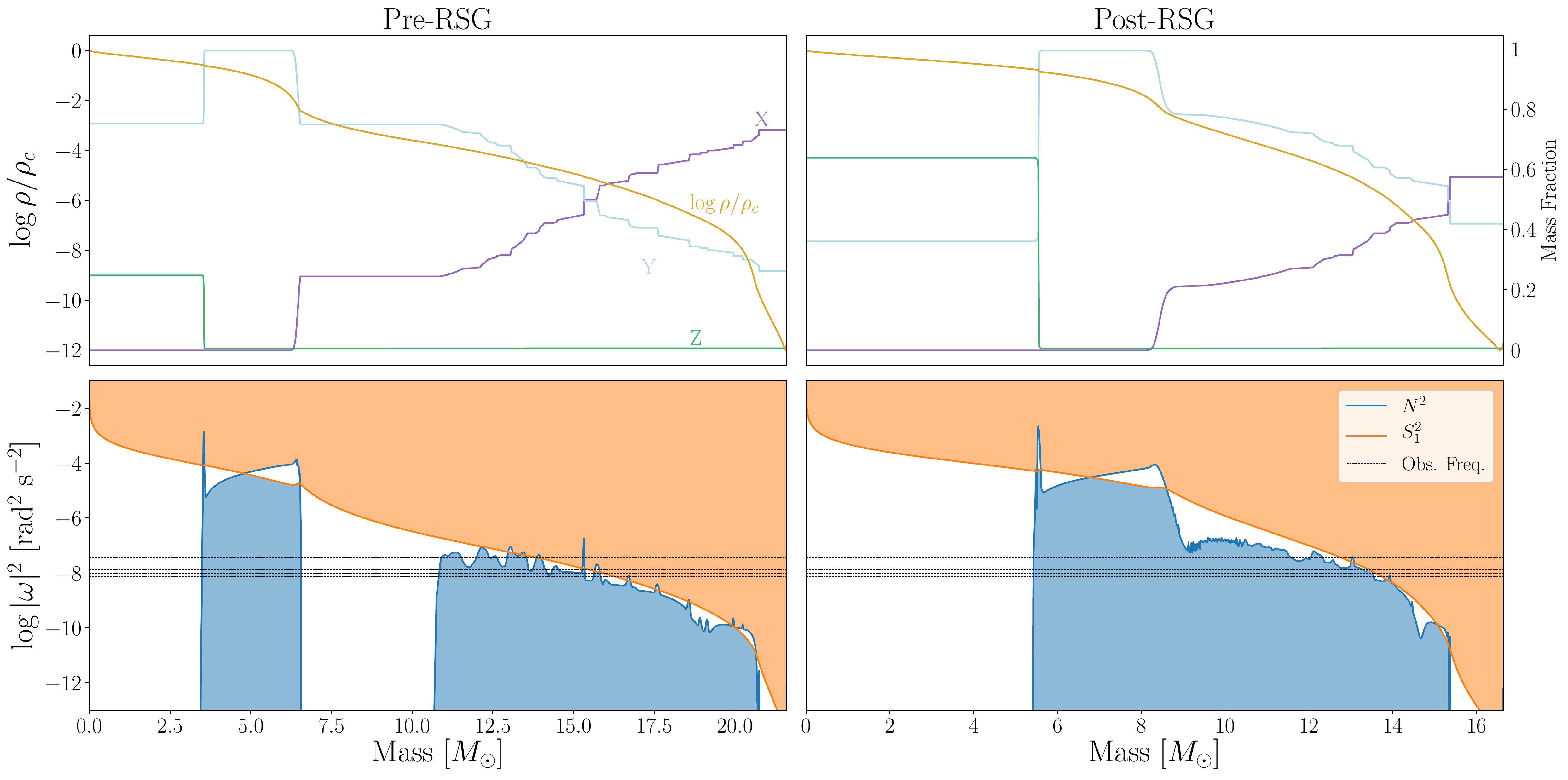}
\caption{{\it Top}: Interior structure of the closest pre-RSG (left) and post-RSG (right) 28 $M_\odot$ MESA models to HD 269953. The dark yellow line shows the density profile, while the purple, light blue, and green lines correspond to the profiles of $X$, $Y$, and $Z$, respectively, as in the right panel of Figure \ref{fig:tracks_hr}. {\it Bottom}: Calculated values of the log of the square of the Brunt-V{\"a}is{\"a}l{\"a} (blue line) and Lamb (orange line) frequencies as a function of the mass coordinate within both MESA models. Here we only show $S_1$ for simplicity. Regions where $\omega < |N|$ and $|\omega| < |S_1|$ (i.e., where {\it g}-modes of a given frequency can propagate) are shaded in blue, and regions where $\omega > |N|$ and $|\omega| > |S_1|$ (i.e., where {\it p}-modes of a given frequency can propagate) are shaded in orange. Frequencies extracted from the \tess lightcurve of HD 269953 are shown as horizontal black dashed lines.} \label{fig:propagation}
\end{figure*}

We show the interior structures of both models as a function of mass coordinate in the top panels of Figure \ref{fig:propagation}; the pre-RSG model is on the left, the post-RSG model is on the right. The density profiles, normalized by the central density $\rho_c$, are shown in dark yellow, while the composition is shown as the profiles of $X$, $Y$, and $Z$ with identical colors as the right panel of Figure \ref{fig:tracks_hr}. We calculate the Lamb and Brunt-V{\"a}is{\"a}l{\"a} frequencies in each model. The bottom panels of Figure \ref{fig:propagation} show the logarithm of $N^2$ (blue) and $S_1^2$ (orange). The frequencies detected in the \tess lightcurve of HD 269953 are shown as horizontal black dashed lines. Two regions exist within the pre-RSG model where {\it g}-modes are able to propagate (shaded in blue), while {\it p}-modes are able to propagate throughout the envelope of the star (shaded in orange). The inner {\it g}-mode cavity corresponds to the region outside of the core that experienced previous H burning, while the outer cavity cavity corresponds to a chemically-stratified outer envelope. In the post-RSG model, the innermost {\it g}-mode cavity has moved outward, and combined with the outer cavity. Sharp features in the stellar structure can be seen in both characteristic frequencies.\footnote{We note that the sharp steps in the composition profiles of the envelopes of both models (and the corresponding spiky behavior of $N^2$ in this region) are due to small discontinuities in the resolution of the MESA model, and are not real features.} 

We stress that this is {\it not} an asteroseismic analysis of the star; we have made no attempt at predicting the excited frequencies in either model, and are by no means identifying the observed frequencies with {\it p-} or {\it g-}modes, let alone more complicated phenomena such as strange modes. Furthermore, the exact treatment of mixing (including semiconvection and thermohaline mixing, which we do not include in our simple models) can have an incredibly strong infuence on the pre-supernova structure of the star \citep[e.g.][]{farmer16}. Indeed, with no reliable interior models of YSGs, we cannot identify modes in order to conduct a full asteroseismic analysis, which may allow us to constrain these physics as well as the evolutionary status of FYPS. However, this rudimentary comparison does illuminate the regions of the stellar structure that the observed pulsations might probe, as well as the drastically different interior structures and pulsational properties seen in pre- and post-RSG models that reside in quite similar regions of the HR diagram. One possible step towards mode identification is to see whether the highest amplitude pulsation frequencies scale with the observed parameters of the stars to ascertain whether the pulsations may be in the acoustic, gravity, or gravito-intertial regime. We searched for correlations between the strongest observed frequencies and $\log T_{eff}$, $\log L/L_\odot$, and $\log R^{-2}$ (as a proxy for $\log g$), but a sample of only five FYPS is insufficient to find any obvious trends. Futhermore, the available spectra are of insufficient resolution to measure $v \sin i$, which are typically less than 10 km s $^{-1}$ in YSGs \citep{barbuy96}. With more FYPS, and higher resolution spectroscopy we may be able to conclusively determine the origin of these pulsations. Again, we emphasize that this demonstration serves only to motivate future work on these stars, and show the potential of FYPS for asteroseismology of YSGs.

\subsection{Evolutionary Status: Leftward vs. Rightward Evolution, and the Red Supergiant Problem}

Examining Figure \ref{fig:tracks_hr}, the FYPS appear to be descended from stars with initial masses of  $M_{FYPS}\gtrsim20 M_\odot$. As discussed above, the maximum observed luminosity of Supernovae II-P progenitors is significantly lower than the maximum luminosity of field RSGs (the red supergiant problem), implying that stars with an initial mass above $M_{max}\approx20 M_\odot$ don't end their lives as red supergiants that explode \citep{kochanek20}. The coincidence between $M_{max}$ and $M_{FYPS}$ is consistent with perhaps the most natural solution to the red supergiant problem: high mass RSGs aren't found as supernova progenitors because they do not explode as RSGs. They instead evolve bluewards on the HR diagram after losing significant mass as RSGs, attaining high enough values of $L/M$ to excite the rich spectrum of observed pulsation modes as predicted in the post-RSG models of \citet{saio13}; FYPS are thus post-RSG objects.\footnote{Note that the MESA models more massive than 18 $M_\odot$ lose enough mass as RSGs to evolve bluewards in the HR diagram. This is not seen in the low-metallicity Geneva models \citep{georgy12,leitherer14}. If FYPS are genuine post-RSGs, this would require an increase in the mass loss rates used in standard stellar models. RSG mass loss rates have only been directly measured in very small numbers of stars in the Galaxy \citep[e.g.][]{mauron11}, and in even fewer LMC RSGs \citep[e.g.][]{vanloon05}. RSG mass loss is also episodic, and while constant mass loss prescriptions can reproduce the makeup of RSG populations \citep{neugent20}, they are still only an approximation to the actual mass loss histories of individual objects.}

While this hypothesis offers a tantalizing solution, putting it on more solid footing would require additional evidence that FYPS are indeed in a post-RSG phase. One route is to search for evidence of strong past mass loss. The TIC contains photometry from the Wide Field Infrared Exlorer (WISE, \citealt{wright10}), in four mid-infrared bands: $W1$ (3.4 $\mu$m), $W2$ (4.6 $\mu$m), $W3$ (12 $\mu$m), and $W4$ (22 $\mu$m). For most stars, the WISE bands are in the Rayleigh-Jeans tail of their spectral energy distribution (SED), and they therefore have colors $\sim0$ in Vega passbands \citep{davenport14}. Deviations from 0 may be attributable to infrared molecular bands, especially in the spectra of red supergiants (Dicenzo \& Levesque, in prep). However, they may also be indicative of circumstellar dust.

We inspected the WISE colors of all stars in our sample. Only the brightest and coolest FYPS, HD 269953, has significantly larger values in each WISE color than nearby stars on the HR diagram. The remaining FYPS are mostly indistinguishable from their fellow stars, though HD 269840 also has a slightly larger $W3-W4$ value than other nearby stars. The three FYPS discussed in \S\ref{subsec:comb} have somewhat {\it smaller} values of $W3-W4$, more consistent with the RSGs than the YSGs. While suggestive that the two most luminous FYPS may have some warm dust contributing to their SEDs, this is not concrete evidence that all five FYPS are in a post-RSG phase. Of course, these mid-infrared measurements are only sensitive to warm dust located relatively close to the star. Perhaps HD 269953 is a less-evolved FYPS with a warm and close CSM, while the CSM of the remaining FYPS have cooled and are undetectable. Probing cooler dust further from the stellar surface would require higher resolution imaging at longer wavelengths \citep[e.g.][]{shenoy16}. If the mechanism powering the pulsations in FYPS also operates at higher metallicity, it would be possible to find a more nearby FYPS in our Galaxy in order to perform such observations. 

It is also important to note that the circumstellar dust produced by RSGs is known to be larger grain than that found in the interstellar medium, and that incorrectly accounting for ``gray" extinction from this dust can lead to underestimates of $L$ in RSGs (e.g. \citealt{massey05,scicluna15,haubois19,levesque20}). If the same is true for the circumstellar dust found around post-RSG YSGs (produced during the RSG phase), it is possible that this could lead to underestimating the $L$ of these stars, which would in turn lead to underestimating stellar mass when comparing the stars to evolutionary tracks on the HR diagram. Quantifying this requires careful observations of these stars' circumstellar environments and dust properties.

One concrete counterargument to the post-RSG hypothesis is found in HD 269110. If HD 269110 is a binary, and the pulsation amplitudes are orbitally modulated, such a binary system would have to be relatively close and eccentric, implying that the two stars have not previously exchanged significant mass. If the primary had previously experienced strong mass-loss as an RSG, the orbit would most likely have circularized instead. However, as discussed above, a companion star with a 91 day orbit would have to be $\sim100 M_\odot$ in order for the semimajor axis of the orbit to be larger than the stellar radius. As rotation is an equally unlikely culprit due to the low surface gravity/critical rotation speeds typical of YSGs, a more thorough characterization of HD 269110 is required.

Ultimately, it is hard to draw any conclusions from such a small sample of pulsators. A total of 341 YSGs identified from \citet{humphreys78}, \citet{sowell90}, \citet{venn95}, \citet{gray01} --- all in the Milky Way --- \citep{neugent10} --- SMC --- and \citep{neugent12_ysg} have entries in the TIC. Assuming our sample of YSGs observed by \tess is representative of the entire population, $\sim5/27=19\%$ of YSGs are FYPS, implying that \tess can detect $\sim63$ FYPS. Even if \tess observes half as many FYPS, we would be able to better characterize the boundaries of the region in the HR diagram in which they reside, and whether the presence/behavior of FYPS is dependent on metallicity. 

\section{Conclusions}\label{sec:conclusion}

Our main results are summarized as follows:

\begin{itemize}
  \item We study the \tess lightcurves of 76 cool supergiants with accurate temperatures and luminosities. For YSGs located in the Large Magellanic Cloud, these lightcurves span a time baseline of a year. We discover that low-frequency stochastic variability is ubiquitous in these stars, and rule out surface convection as the underlying cause for all but the red supergiants. This implies that this variability, also observed in main sequence O stars, is a constant feature of massive stars throughout their lifetimes.
  \item After removing the contribution of this background variability from the periodograms of the stars in our sample, we find two regions in the HR diagram with pulsating stars. Four of these stars are candidate $\alpha$ Cygni variables, of which three are newly identified as such. The remaining five pulsating stars are clustered in a region of the HR diagram not previously identified as a region of instability. 
  \item We rule out binarity, spurious signals, and chance alignment with nearby stars causing us to mistakenly find these pulsators in the same part of the HR diagram by chance, and conclude that these five stars comprise a real class, that we dub Fast Yellow Pulsating Supergiants (FYPS).
  \item We extract pulsation frequencies from the FYPS lightcurve using a new procedure to account for the stochastic background, search for harmonics and frequency combinations from the extracted frequencies, and calculate the Weighted Wavelet Z-transform (WWZ) of the lightcurves to study the time-dependent behavior of these frequencies 
  \item In HD 269953 and HD 269840 (and perhaps HD 269902), the lowest frequency strong periodogram peak extracted is coincident with a broad ridge of power in the WWZ that rapidly switches frequencies and amplitudes on $\sim$day timescales.
  \item HD 268687, HD 269840, and HD 269902 show a broad comb of frequencies, with most extracted peaks being close to but not exactly in a harmonic chain.
  \item One of the frequencies found in HD 269110 is split into a triplet, and the WWZ of the lightcurve at that frequency is modulated on the same timescale as the difference between peaks in the triplet. We are unable to determine whether or not this is due to rotation, binary effects, or some other cause.
  \item We introduce the possibility that FYPS are post-RSG objects that have lost enough mass as RSGs to attain high luminosity-to-mass ratios, and excite pulsations \citep[as in][]{saio13}. This possibility is bolstered by the coincidence between the lowest estimated mass of the FYPS and the highest mass RSG progenitors. While we are unable to concretely determine the exact evolutionary status of the FYPS, future work to determine whether or not they are genuine post-RSG objects is of extreme importance, both theoretically by modeling their pulsation frequencies, and observationally by finding more FYPS. Regardless of the evolutionary status of FYPS, their pulsational properties will be of great use in the study of the interiors of evolved massive stars.
\end{itemize}
\vspace{-20pt}
\acknowledgments
The authors acknowledge that the work presented was largely conducted on the traditional land of the first people of Seattle, the Duwamish People past and present and honor with gratitude the land itself and the Duwamish Tribe. This research  was  supported  by  NSF  grant AST 1714285 and a Cottrell Scholar Award from the Research Corporation for Scientific Advancement granted to EML. This work has been carried out in the framework of the PlanetS National Centre of Competence in Research (NCCR) supported by the Swiss National Science Foundation (SNSF). 

The authors thank C. Georgy for sharing the $Z=0.006$ Geneva evolutionary tracks.

This research has made use of the VizieR catalogue access tool, CDS, Strasbourg, France (DOI: 10.26093/cds/vizier). The original description of the VizieR service was published in A\&AS 143, 23. This research has made use of the SIMBAD database, operated at CDS, Strasbourg, France.

The data described here may be obtained from the MAST archive at
\dataset[doi:10.17909/T9RP4V]{https://dx.doi.org/10.17909/T9RP4V}.

This work made use of the following software and facilities:

\vspace{5mm}

\facility{The {\it Transiting Exoplanet Survey Satellite} (\tess, \citealt{ricker15})}

\software{Astropy v3.1.2 \citep{astropy13,astropy18}, Matplotlib v3.1.1 \citep{Hunter:2007}, MESA r12778 \citep{paxton11,paxton13,paxton15,paxton18,paxton19}, NumPy v1.17.2 \citep{numpy:2011}, Pandas v0.25.1 \citep{pandas:2010},  Python 3.7.4, Scipy v1.3.1 \citep{scipy:2001}, Astroquery \citep{astroquery}}

\newpage
\bibliography{bib}
\bibliographystyle{aasjournal}

\appendix
\restartappendixnumbering
\section{Frequencies found via prewhitening}\label{app:A}

The following tables contain the list of unique frequencies (separated by $1.5/T$) found in the lightcurves of the FYPS by the prewhitening procedure described in \S\ref{subsec:prewhitening}. Frequencies that are exact harmonics of other frequencies are indicated. Combination frequencies recovered are noted in the comments of each table. However, these are only the ``exact'' harmonics and combination frequencies recovered to within the precision of the observed frequencies. As discussed in \S\ref{subsec:comb}, frequencies with spacings that are close to but not exact harmonics and combinations are recovered in HD 268687, HD 269840, and HD 269902; in these stars, only $f_4$, $f_4$ and $f_6$, and $f_2$ respectively don't belong to these sequences of near-harmonics.

\begin{deluxetable*}{lcccccccc}
\tabletypesize{\footnotesize}
\tablecaption{Unique frequencies, amplitudes, phases, and formal errors for HD 269953 found via prewhitening. For each frequency, we specify the SNR as defined in text, and the height of the associated peak in the RPS at that stage of prewhitening.\label{tab:HD 269953_freqs}}
\tablehead{\colhead{Frequency} & \colhead{$f_j$} & \colhead{$\epsilon(f_j)$} & \colhead{$A_j$} & \colhead{$\epsilon(A_j)$} & \colhead{$\phi_j$} &  \colhead{$\epsilon(\phi_j)$} & \colhead{SNR} & \colhead{RPS Peak Height}\\
\colhead{} & \colhead{[day$^{-1}$]} & \colhead{[day$^{-1}$]} & \colhead{[ppt]} & \colhead{[ppt]} & \colhead{[radians]} & \colhead{[radians]} & \colhead{} & \colhead{} } 
\startdata
$f_{0}$& $1.59347960$ & $0.00002335$ & $0.17835463$ & $0.00269747$ & $2.2998$ & $0.0151$ & 53.8321 & 1187.4096 \\ 
$f_{1}^*$& $2.67052158$ & $0.00009247$ & $0.04452312$ & $0.00266624$ & $0.2701$ & $0.0599$ & 32.9642 & 150.6893 \\ 
$f_{2}$& $1.33523329$ & $0.00006790$ & $0.06058007$ & $0.00266395$ & $-0.5273$ & $0.0440$ & 20.1564 & 133.2001 \\ 
$f_{3}$& $1.17424693$ & $0.00006774$ & $0.06062141$ & $0.00265962$ & $0.2279$ & $0.0439$ & 29.5097 & 118.7172 \\ 
\enddata 
\tablecomments{$*$: harmonics of $f_{2}$. } 
\end{deluxetable*} 

\begin{deluxetable*}{lcccccccc}
\tabletypesize{\footnotesize}
\tablecaption{Unique frequencies, amplitudes, phases, and formal errors for HD 269110 found via prewhitening. For each frequency, we specify the SNR as defined in text, and the height of the associated peak in the RPS at that stage of prewhitening.\label{tab:HD 269110_freqs}}
\tablehead{\colhead{Frequency} & \colhead{$f_j$} & \colhead{$\epsilon(f_j)$} & \colhead{$A_j$} & \colhead{$\epsilon(A_j)$} & \colhead{$\phi_j$} &  \colhead{$\epsilon(\phi_j)$} & \colhead{SNR} & \colhead{RPS Peak Height}\\
\colhead{} & \colhead{[day$^{-1}$]} & \colhead{[day$^{-1}$]} & \colhead{[ppt]} & \colhead{[ppt]} & \colhead{[radians]} & \colhead{[radians]} & \colhead{} & \colhead{} } 
\startdata
$f_{0}$& $0.55280981$ & $0.00003493$ & $0.16642430$ & $0.00376526$ & $-1.7283$ & $0.0226$ & 19.2635 & 173.6190 \\ 
$f_{1}$& $1.76353979$ & $0.00013526$ & $0.04277167$ & $0.00374665$ & $1.0248$ & $0.0876$ & 28.0123 & 59.3401 \\ 
$f_{2}$& $0.54185885$ & $0.00007463$ & $0.07747331$ & $0.00374458$ & $-1.2620$ & $0.0483$ & 14.7085 & 40.2324 \\ 
$f_{3}$& $0.56377424$ & $0.00009201$ & $0.06277635$ & $0.00374066$ & $-2.2209$ & $0.0596$ & 7.5121 & 29.0397 \\ 
\enddata 
\end{deluxetable*} 

\begin{deluxetable*}{lcccccccc}
\tabletypesize{\footnotesize}
\tablecaption{Unique frequencies, amplitudes, phases, and formal errors for HD 268687 found via prewhitening. For each frequency, we specify the SNR as defined in text, and the height of the associated peak in the RPS at that stage of prewhitening.\label{tab:HD 268687_freqs}}
\tablehead{\colhead{Frequency} & \colhead{$f_j$} & \colhead{$\epsilon(f_j)$} & \colhead{$A_j$} & \colhead{$\epsilon(A_j)$} & \colhead{$\phi_j$} &  \colhead{$\epsilon(\phi_j)$} & \colhead{SNR} & \colhead{RPS Peak Height}\\
\colhead{} & \colhead{[day$^{-1}$]} & \colhead{[day$^{-1}$]} & \colhead{[ppt]} & \colhead{[ppt]} & \colhead{[radians]} & \colhead{[radians]} & \colhead{} & \colhead{} } 
\startdata
$f_{0}$& $2.76530509$ & $0.00006476$ & $0.18020143$ & $0.00755737$ & $-2.9889$ & $0.0419$ & 63.1039 & 268.6123 \\ 
$f_{1}^*$& $3.68693149$ & $0.00010893$ & $0.10696509$ & $0.00754615$ & $2.7915$ & $0.0705$ & 86.6110 & 171.2117 \\ 
$f_{2}$& $1.84352713$ & $0.00005631$ & $0.20680066$ & $0.00754233$ & $-2.5874$ & $0.0365$ & 44.4964 & 148.7754 \\ 
$f_{3}$& $4.60860763$ & $0.00018083$ & $0.06428032$ & $0.00752813$ & $2.2392$ & $0.1171$ & 57.2225 & 88.0887 \\ 
$f_{4}$& $1.34234726$ & $0.00004934$ & $0.23553062$ & $0.00752674$ & $3.0867$ & $0.0320$ & 50.1173 & 87.3329 \\ 
$f_{5}$& $0.92187126$ & $0.00004733$ & $0.24496247$ & $0.00750823$ & $-2.2787$ & $0.0307$ & 8.2893 & 33.9434 \\ 
\enddata 
\tablecomments{$*$: harmonics of $f_{2}$. $f_3 = f_1 + f_5$. $f_0 = f_2 + f_5$. } 
\end{deluxetable*} 

\begin{deluxetable*}{lcccccccc}
\tabletypesize{\footnotesize}
\tablecaption{Unique frequencies, amplitudes, phases, and formal errors for HD 269840 found via prewhitening. For each frequency, we specify the SNR as defined in text, and the height of the associated peak in the RPS at that stage of prewhitening.\label{tab:HD 269840_freqs}}
\tablehead{\colhead{Frequency} & \colhead{$f_j$} & \colhead{$\epsilon(f_j)$} & \colhead{$A_j$} & \colhead{$\epsilon(A_j)$} & \colhead{$\phi_j$} &  \colhead{$\epsilon(\phi_j)$} & \colhead{SNR} & \colhead{RPS Peak Height}\\
\colhead{} & \colhead{[day$^{-1}$]} & \colhead{[day$^{-1}$]} & \colhead{[ppt]} & \colhead{[ppt]} & \colhead{[radians]} & \colhead{[radians]} & \colhead{} & \colhead{} } 
\startdata
$f_{0}$& $2.26965893$ & $0.00005279$ & $0.19821747$ & $0.00463827$ & $-1.0031$ & $0.0234$ & 60.4289 & 490.7758 \\ 
$f_{1}$& $3.40445242$ & $0.00007777$ & $0.13350616$ & $0.00460197$ & $2.2648$ & $0.0345$ & 45.3612 & 460.6608 \\ 
$f_{2}$& $1.13468853$ & $0.00003708$ & $0.27915368$ & $0.00458731$ & $1.8206$ & $0.0164$ & 41.2142 & 332.8405 \\ 
$f_{3}$& $2.83703781$ & $0.00020544$ & $0.04966845$ & $0.00452268$ & $2.2047$ & $0.0911$ & 33.1261 & 61.2621 \\ 
$f_{4}$& $0.71676916$ & $0.00007334$ & $0.13905708$ & $0.00451996$ & $0.5576$ & $0.0325$ & 18.0708 & 57.1411 \\ 
$f_{5}$& $3.97206611$ & $0.00024415$ & $0.04160807$ & $0.00450260$ & $-0.9063$ & $0.1082$ & 32.0854 & 55.6234 \\ 
$f_{6}$& $1.38512974$ & $0.00013130$ & $0.07734552$ & $0.00450113$ & $2.4916$ & $0.0582$ & 24.1347 & 52.1294 \\ 
\enddata 
\end{deluxetable*} 

\begin{deluxetable*}{lcccccccc}
\tabletypesize{\footnotesize}
\tablecaption{Unique frequencies, amplitudes, phases, and formal errors for HD 269902 found via prewhitening. For each frequency, we specify the SNR as defined in text, and the height of the associated peak in the RPS at that stage of prewhitening.\label{tab:HD 269902_freqs}}
\tablehead{\colhead{Frequency} & \colhead{$f_j$} & \colhead{$\epsilon(f_j)$} & \colhead{$A_j$} & \colhead{$\epsilon(A_j)$} & \colhead{$\phi_j$} &  \colhead{$\epsilon(\phi_j)$} & \colhead{SNR} & \colhead{RPS Peak Height}\\
\colhead{} & \colhead{[day$^{-1}$]} & \colhead{[day$^{-1}$]} & \colhead{[ppt]} & \colhead{[ppt]} & \colhead{[radians]} & \colhead{[radians]} & \colhead{} & \colhead{} } 
\startdata
$f_{0}$& $2.90337860$ & $0.00006011$ & $0.12717671$ & $0.00455148$ & $-1.2740$ & $0.0358$ & 62.4969 & 333.8202 \\ 
$f_{1}$& $1.45175166$ & $0.00003744$ & $0.20368400$ & $0.00454088$ & $-1.4444$ & $0.0223$ & 53.9124 & 307.3930 \\ 
$f_{2}$& $0.08163766$ & $0.00002411$ & $0.31421154$ & $0.00451126$ & $-2.6455$ & $0.0144$ & 5.5540 & 21.4397 \\ 
\enddata 
\end{deluxetable*}

\end{document}